\documentclass[twocolumn,english,nofootinbib]{revtex4}
\usepackage[T1]{fontenc}
\usepackage[latin9]{inputenc}
\usepackage{color}
\usepackage{amsmath}
\usepackage{amssymb}
\usepackage{graphicx}

\makeatletter

\providecommand{\tabularnewline}{\\}

\@ifundefined{textcolor}{}
{%
 \definecolor{BLACK}{gray}{0}
 \definecolor{WHITE}{gray}{1}
 \definecolor{RED}{rgb}{1,0,0}
 \definecolor{GREEN}{rgb}{0,1,0}
 \definecolor{BLUE}{rgb}{0,0,1}
 \definecolor{CYAN}{cmyk}{1,0,0,0}
 \definecolor{MAGENTA}{cmyk}{0,1,0,0}
 \definecolor{YELLOW}{cmyk}{0,0,1,0}
}

\makeatother

\usepackage{babel}
\begin{document}

\title{Gravitational wave emission and spindown of young pulsars}

\author{Mark G. Alford and Kai Schwenzer}

\address{Department of Physics, Washington University, St. Louis, Missouri,
63130, USA}
\begin{abstract}
The rotation frequencies of young pulsars are systematically below
their theoretical Kepler limit. R-modes have been suggested as a possible
explanation for this observation. With the help of semi-analytic expressions
that make it possible to assess the uncertainties of the r-mode scenario
due to the impact of uncertainties in underlying microphysics, we
perform a quantitative analysis of the spin-down and the emitted gravitational
waves of young pulsars. We find that the frequency to which r-modes
spin down a young neutron star is surprisingly insensitive both to
the microscopic details and the saturation amplitude. Comparing our
result to astrophysical data, we show that for a range of sufficiently
large saturation amplitudes r-modes provide a viable spindown scenario
and that all observed young pulsars are very likely already outside
the r-mode instability region. Therefore the most promising sources
for gravitational wave detection are unobserved neutron stars associated
with recent supernovae, and we find that advanced LIGO should be
able to see several of them. We find the remarkable result that the
gravitational wave strain amplitude is completely independent of both
the r-mode saturation amplitude and the microphysics, and depends
on the saturation mechanism only within some tens of per cent. However,
the gravitational wave frequency depends on the amplitude and we provide
the required expected timing parameter ranges to look for promising
sources in future searches.
\end{abstract}
\maketitle

\section{Introduction}

A neutron star is a complex system whose behavior is influenced by
physics on scales from the Fermi scale to its size and whose description
involves all forces of nature. This holds in particular when dynamical
aspects involving mechanical oscillations coupled to radiation fields
are considered and a description should thereby intricately depend
on all these details. Yet, the surprising finding presented in this
work is that for the relevant observables analytic expressions can
be derived that are strikingly insensitive to these complicated microscopic
as well as macroscopic details. 

The rotation frequencies of pulsars, as well as their time derivatives,
are among the most precise observations in physics. These frequencies
change over time due to magnetic braking and other possible spindown
mechanisms like gravitational wave emission due to deformations or
oscillation modes of the star. In particular the unstable r-modes
\cite{Andersson:1997xt,Andersson:2000mf} are interesting because
they must be stabilized by a viscous dissipation mechanism and thereby
can probe the microphysics deep inside the star. It has already been
shown \cite{Alford:2010fd,Alford:2011pi} that important static aspects
of the instability regions of r-modes are very insensitive to quantitative
microscopic details but do depend on qualitative differences of the
various possible forms of dense matter. Here we extend this study
to the dynamical r-mode evolution of neutron stars and show that in
this case the insensitivity to unknown input parameters is even more
pronounced. In particular we derive semi-analytic expressions for
the final frequency of the spindown evolution and the spindown time
of young neutron stars \cite{Andersson:1998ze} and show that these
quantities are likewise insensitive to the particular r-mode saturation
mechanism.

Magnetic dipole radiation is considered as a standard mechanism for
the spindown of pulsars and is used to determine an approximate characteristic
spindown age \cite{1983bhwd.book.....S}. In contrast, in this work
we will instead only consider the spindown torque due to r-modes,
which is only non-vanishing as long as the star is within the unstable
region. Our results therefore provide upper limits on the actual frequencies
of observed pulsars and should apply as long as the r-mode spindown
dominates. However, they might not be applicable to magnetars where
this is not guaranteed. A more detailed analysis which includes both
electromagnetic and gravitational radiation \cite{Palomba:1999su,Staff:2011zn}
and thereby allows to make contact to pulsar ages will be given elsewhere.

The fastest young pulsar observed so far is PSR J0537-6910 with a
rotational frequency of $f\!\approx\!62$ Hz, which has been associated
to the remnant N157B of a supernova that exploded in the Large Magellanic
Cloud \cite{Marshall:1998}. We find that this spin frequency is just
below the final frequency due to pure r-mode emission of a neutron
star with standard modified Urca beta equilibration processes, so
that r-mode emission could indeed provide a quantitative explanation
for the low spin frequencies of young neutron stars. If the r-mode
scenario is realized, the fact that all observed stars are below the
r-mode bound also suggests that r-mode spindown should be fast, i.e.\ at
least comparable to magnetic spindown, which would require a high
r-mode saturation amplitude $\alpha\!\gtrsim\!0.01$ in young neutron
stars.

For standard damping, i.e.\ in the absence of hypothetical strong
additional sources of dissipation, r-modes of neutron stars are an
important source of gravitational waves and provide an interesting
possibility for their detection since r-modes can radiate over long
times. Our results on the spindown show that if r-modes explain young
neutron star spins, that implies a large saturation amplitude which
in turn says that the window of opportunity for direct observation
is brief. Promising sources are therefore very young and so far undetected
rapidly spinning neutron stars. Previously analytic estimates have
been given for the gravitational wave strain of r-modes of young stars
\cite{Owen:1998xg}. However, this work did not take into account
the limited observation time in practical gravitational wave searches
and found unrealistically large signal-to-noise ratios. Using the
semi-analytic expression for the r-mode evolution we perform a comprehensive
analysis of the gravitational wave emission. We find the remarkable
result that the gravitational wave emission from r-modes in young
pulsars is \emph{independent} of the saturation amplitude and even
shows hardly any dependence on the saturation mechanism. This allows
us to give surprisingly definite predictions for the gravitational
wave signal despite our limited understanding of these complicated
systems. We compare our results to the detector sensitivities for
realistic searches that take into account our ignorance of the detailed
properties of the source and find that forthcoming second generation
detectors like advanced LIGO \cite{Harry:2010zz,Andersson:2009yt}
are sensitive enough to see several potential sources.

\section{Material and star properties}

As noted before a neutron star is a complex system and the description
of its spindown evolution requires an understanding of various different
aspects. The staring point is the static star configuration which
is determined by gravity as described by the general relativistic
Oppenheimer-Volkoff (OV) equations \cite{Oppenheimer:1939ne,Tolman:1939jz}.
The latter require the equation of state of charge-neutral dense matter
in beta equilibrium as an ingredient which is determined by microscopic
strong interactions. Both the rotation and the oscillation of the
star are modeled as perturbations of the static respectively the rotating
star configuration \cite{Lindblom:1999yk}. For simplicity both are
standardly obtained from the Newtonian %
\footnote{This Newtonian approximation, as well as the approximation to use
the Eulerian density fluctuation instead of the Lagrangian fluctuation
to evaluate eq.~(\ref{eq:delta-Sigma}) below (see \cite{Lindblom:1999yk}
for details), is only used for numerical estimates. The general semi-analytic
expressions derived in this work are - with appropriate parameter
values - valid in the fully relativistic case and do not rely on such
approximations. These results will show that the impact of these approximations
on the final observables is small.%
} Euler-equation of a non-viscous fluid. The classical r-modes with
$l=m$ are given by the Eulerian velocity perturbations

\begin{equation}
\delta\vec{v}=\alpha R\Omega\left(\frac{r}{R}\right)^{m}\vec{Y}_{mm}^{B}\!\left(\theta,\varphi\right)\mathrm{e}^{i\omega t}\:,\label{eq:r-mode}
\end{equation}
where $\alpha$ is a dimensionless amplitude parameter, $R$ is the
star radius, $\Omega$ and $\omega$ are rotational and mode angular
velocity, respectively, and $\vec{Y}^{B}$ is the magnetic vector
spherical harmonic with magnetic quantum number $m$.

To describe the dynamical evolution requires microscopic material
properties of the dense matter within the star. Here we follow the
method that was developed previously in \cite{Alford:2010fd} and
\cite{Alford:2011pi} to obtain semi-analytic results for the boundary
of the r-mode instability region as well as for the saturation amplitude
in case suprathermal bulk viscosity saturates the mode. The basic
idea is that the microscopic material properties that are relevant
for the pulsar evolution have - at least over the range required to
describe the relevant aspects - simple power law dependencies on the
temperature $T$, the oscillation frequency $\omega$ and the density
oscillation amplitude $\Delta n/\bar{n}$. The required quantities
for the star evolution are the shear viscosity $\eta$, the bulk viscosity
$\zeta$, the specific heat $c_{V}$ and the neutrino emissivity $\epsilon$.
Both the the bulk viscosity and the neutrino emissivity are determined
by a weak equilibration process whose rate in a degenerate system
has the form $\Gamma^{(\leftrightarrow)}\!=\!-\tilde{\Gamma}T^{\delta}\mu_{\Delta}$,
where $\mu_{\Delta}$ is the chemical potential difference that is
driven out of equilibrium. As discussed below, for our present analysis
the suprathermal amplitude dependence of the bulk viscosity and the
neutrino emissivity \cite{Alford:2010gw,Alford:2011df} does not have
a significant impact and we will discuss their impact in other situations
in more detail in a future work. In the subthermal regime these quantities
can generally be parameterized as

\begin{align}
 & \eta=\tilde{\eta}T^{-\sigma}\quad,\quad\zeta\approx\frac{C^{2}\tilde{\Gamma}T^{\delta}}{\omega^{2}+\left(B\tilde{\Gamma}T^{\delta}\right)^{2}}\xrightarrow{}\frac{C^{2}\tilde{\Gamma}T^{\delta}}{\omega^{2}}\:,\label{eq:viscosity-parametrization}\\
 & c_{V}=\tilde{c}_{V}T^{\upsilon}\quad,\quad\epsilon\approx\tilde{\epsilon}T^{\theta}\:,\label{eq:thermal-parametrization}
\end{align}
 where $B$ and $C$ are non-equilibrium susceptibilities. With
such power law behavior it is clear that physical results should depend
far more sensitively on the exponents $\sigma$, $\delta$, $\upsilon$
and $\theta$ than on the prefactor functions $\tilde{\eta}$, $\tilde{c}_{V}$,
$\tilde{\epsilon}$, $\cdots$. The important point is that the exponents
are the same for a given phase of dense matter, like e.g.\ hadronic
matter with modified Urca reactions \cite{Friman:1978zq}, and quantitative
details due to the unknown equation of state or the interactions are
only reflected in the prefactors. Strikingly we will see below that
the quantitative dependence on these prefactors can be even far more
insensitive than this general argument suggests. Different phases
of dense matter, however, can feature qualitatively different low
energy degrees of freedom resulting in different exponents and thereby
can lead to a qualitatively very different star evolution. 

The above material properties are local quantities, but all that enters
the evolution equations below are quantities that are averaged over
the entire star. These are the power radiated in gravitational waves
$P_{G}$, the dissipated power $P_{S}$ and $P_{B}$ due to shear
and bulk viscosity, the specific heat of the star $C_{V}$, the total
neutrino luminosity $L_{\nu}$ and the moment of inertia $I$ \cite{Owen:1998xg,Alford:2010fd}

\begin{align}
P_{G} & =\frac{32\pi\left(m\!-\!1\right)^{2m}\left(m\!+\!2\right)^{\!2m\!+\!2}}{\left(\left(2m\!+\!1\right)!!\right)^{2}\left(m\!+\!1\right)^{\!2m\!+\!2}}\tilde{J_{m}}^{2}\! GM^{2}\! R^{2m\!+\!2}\!\alpha^{2}\!\Omega^{2m\!+\!4},\label{eq:gravitational-power}\\
P_{S} & =-\frac{\left(m-1\right)\left(2m+1\right)\tilde{S}_{m}\Lambda_{{\rm QCD}}^{3+\sigma}R^{3}\alpha^{2}\Omega^{2}}{T^{\sigma}}\:,\label{eq:shear-viscosity-dissipation}\\
P_{B} & =-\frac{16m}{\left(2m+3\right)\left(m+1\right)^{5}\kappa^{2}}\frac{\tilde{V}_{m}\Lambda_{{\rm QCD}}^{9-\delta}R^{7}\alpha^{2}\Omega^{4}T^{\delta}}{\Lambda_{{\rm EW}}^{4}}\:,\label{eq:bulk-vicosity-dissipation}\\
C_{V} & =4\pi\Lambda_{{\rm QCD}}^{3-\upsilon}R^{3}\tilde{C}_{V}T^{\upsilon}\:,\label{eq:heat-capacity}\\
L_{\nu} & =\frac{4\pi R^{3}\Lambda_{{\rm QCD}}^{9-\theta}\tilde{L}}{\Lambda_{{\rm EW}}^{4}}T^{\theta}\:,\label{eq:neutrino-luminosity}\\
I & =\tilde{I}MR^{2}\:.\label{eq:momentum-of-inertia}
\end{align}
where $\Lambda_{{\rm QCD}}$ and $\Lambda_{{\rm EW}}$ are characteristic
strong and electroweak scales introduced to make these quantities
dimensionless. With the energy of the r-mode

\begin{equation}
E_{m}=\frac{1}{2}\alpha^{2}\Omega^{2}MR^{2}\tilde{J}\label{eq:mode-energy}
\end{equation}
one can define characteristic time scales for the r-mode growth $\tau_{G}$
and the viscous damping $\tau_{S}$ and $\tau_{B}$ via

\begin{equation}
\frac{1}{\tau_{i}}=-\frac{P_{i}}{2E_{m}}\:.\label{eq:time-scales}
\end{equation}
 so $\tau_{G}$ is negative and $\tau_{S}$ and $\tau_{B}$ are positive.
The above quantities are given in terms of parameters that involve
the integration over the star or alternatively over the layer(s) ranging
from $R_{i}$ to $R_{o}$ that contribute(s) dominantly to this quantity.
These are defined in table \ref{tab:Integral-parameters.}.

\begin{table}
\begin{tabular}{|c|c|}
\hline 
parameter of the ... & integral expression\tabularnewline
\hline 
moment of inertia & $\tilde{I}\equiv\frac{8\pi}{3MR^{2}}\int_{0}^{R}dr\, r^{4}\rho$\tabularnewline
\hline 
radiated power & $\tilde{J_{m}}\equiv\frac{1}{MR^{2m}}\int_{0}^{R}dr\, r^{2m+2}\rho$\tabularnewline
\hline 
shear dissipated power & $\tilde{S}_{m}\equiv\frac{1}{R^{2m+1}\Lambda_{{\rm QCD}}^{3+\sigma}}\int_{R_{i}}^{R_{o}}dr\, r^{2m}\tilde{\eta}$\tabularnewline
\hline 
bulk dissipated power & $\tilde{V}_{m}\equiv\frac{\Lambda_{{\rm EW}}^{4}}{R^{3}\Lambda_{{\rm QCD}}^{9-\delta}}\int_{R_{i}}^{R_{o}}\! dr\, r^{2}A^{2}C^{2}\tilde{\Gamma}\,\left(\delta\!\Sigma_{m}\right)^{2}$\tabularnewline
\hline 
specific heat & $\tilde{C}_{V}\equiv\frac{1}{R^{3}\Lambda_{{\rm QCD}}^{3-\upsilon}}\int_{R_{i}}^{R_{o}}dr\, r^{2}\tilde{c}_{V}$\tabularnewline
\hline 
neutrino luminosity & $\tilde{L}\equiv\frac{\Lambda_{{\rm EW}}^{4}}{R^{3}\Lambda_{{\rm QCD}}^{9-\theta}}\int_{R_{i}}^{R_{o}}dr\, r^{2}\tilde{\epsilon}$\tabularnewline
\hline 
\end{tabular}\caption{\label{tab:Integral-parameters.}Radial integral parameters encoding
the complete information on the star's interior - i.e.\ on the microphysical
transport properties, the equation of state and the star's density
profile.}
\end{table}
All quantities appearing in these integrals, like the energy density
$\rho$, the inverse squared speed of sound $A$, the non-equilibrium
susceptibility $C$ and the r-mode density oscillation

\begin{align}
\delta\Sigma & \equiv\frac{m+1}{2\alpha AR^{2}\Omega^{3}}\sqrt{\frac{\left(m\!+\!1\right)^{3}\left(2m\!+\!3\right)}{4m}}\left(\int d\Omega\left|\vec{\nabla}\cdot\delta\vec{v}\right|^{2}\right)^{\frac{1}{2}}\label{eq:delta-Sigma}
\end{align}
depend on the position within the star via their density dependence.
These few constants encode the complete information on the physics
inside the star and are sufficient to calculate the star evolution.

The dimensionless parameters $\tilde{I}$ and $\tilde{J}$, arising
in the moment of inertia and the canonical energy and angular momentum
of the mode, involve integrals over the energy density in which the
outer parts are strongly weighted by high powers of $r$. They are
normalized by the mass and appropriate powers of the radius so that
the range these constants can vary over is independent of the mass.
The mass by which these constants are divided is given by an analogous
integral with a smaller power of $r$. Consequently these constants
are large for stars whose mass is strongly spread out but small for
stars where the mass is concentrated close to the center. Since the
energy density is a monotonically decreasing function, it is clear
that $\tilde{I}$ and $\tilde{J}$ are bounded from above by the values
for a constant density star \cite{Alford:2010fd}. Similarly, lower
bounds can be obtained by noting that in a star the mass cannot be
arbitrarily concentrated. A limit for the energy density at a given
radius is obtained by the constraint that the matter within this radius
has to be stable against gravitational collapse. The corresponding
bound for the energy density is given by $\rho\!\left(r\right)<1/(8\pi Gr^{2})$
which can be integrated to obtain lower bounds on $\tilde{I}$ and
$\tilde{J}$. Combining these rigorous limits we find that for any
compact star these parameters are narrowly bounded within roughly
a factor of two

\begin{align}
0.22\approx\frac{2}{9} & \leq\tilde{I}\leq\frac{2}{5}=0.4\:,\label{eq:I-tilde-bounds}\\
1.59\cdot10^{-2}\approx\frac{1}{20\pi} & \leq\tilde{J}\leq\frac{3}{28\pi}\approx3.41\cdot10^{-2}\:.\label{eq:J-tilde-bounds}
\end{align}
However, for a neutron star with a crust the energy density vanishes
at the surface and therefore the upper bounds should even clearly
overestimate the realistic range. Assuming expected mass and radius
ranges of a neutron star $1\, M_{\odot}\lesssim M\lesssim2.5\, M_{\odot}$
and $10\,{\rm km}\lesssim R\lesssim15\,{\rm km}$, the moment of inertia
eq. (\ref{eq:momentum-of-inertia}) is therefore uncertain within
at most an order of magnitude

\begin{equation}
4.4\times10^{44}\frac{{\rm g}}{{\rm cm}^{2}}\lesssim I\lesssim4.5\times10^{45}\frac{{\rm g}}{{\rm cm}^{2}}\:.\label{eq:moment-of-inertia-bounds}
\end{equation}
The uncertainty on the other parameters arises both from the microscopic
quantities eqs.~(\ref{eq:viscosity-parametrization}) and (\ref{eq:thermal-parametrization})
as well as from the particular (baryon) density profile $n\!\left(r\right)$
which in turn depends both on the equation of state and the particular
solution of the OV equations (parameterized e.g.\ by the star's mass).
All parameters defined above in tab. \ref{tab:Integral-parameters.}
are given for different stars in tab.~\ref{tab:parameter-values}.
For illustration of our semi-analytic results below we consider different
stars with an APR equation of state \cite{Akmal:1998cf}. In these
stars we assume that the core of the star dominates the relevant quantities
and neglect possible contributions from the crust. For the maximum
mass APR star the density is high enough that direct Urca processes
are kinematically allowed within an inner core%
\footnote{The density becomes large enough at the center so that the direct
Urca channel opens for stars with masses slightly above $2\, M_{\odot}$.%
}. 
\begin{table*}
\begin{tabular}{|c|c|c|c|c|c|c|c|c|c|c|c|c|c|c|}
\hline 
neutron star & shell & $R\left[km\right]$ & $\Omega_{K}\left[Hz\right]$ & $\tilde{I}$ & $\tilde{J}$ & $Q$ & $\tilde{S}$ & $\tilde{V}$ & $\tilde{C}_{V}$ & $\tilde{L}$ & $\sigma$ & $\delta$ & $\upsilon$ & $\theta$\tabularnewline
\hline 
APR $1.4\, M_{\odot}$ & core & $11.5$ & $6020$ & $0.283$ & $1.81\times10^{-2}$ & $0.096$ & $7.68\times10^{-5}$ & $1.31\times10^{-3}$ & $2.36\times10^{-2}$ & $1.91\times10^{-2}$ & $\frac{5}{3}$ & $6$ & $1$ & $8$\tabularnewline
\cline{1-1} \cline{3-11} 
APR $2.0\, M_{\odot}$ &  & $11.0$ & $7670$ & $0.300$ & $2.05\times10^{-2}$ & $0.102$ & $2.25\times10^{-4}$ & $1.16\times10^{-3}$ & $2.64\times10^{-2}$ & $1.69\times10^{-2}$ &  &  &  & \tabularnewline
\cline{1-11} 
APR $2.21\, M_{\odot}$ & m.U. core & $10.0$ & $9310$ & $0.295$ & $2.02\times10^{-2}$ & $0.103$ & $5.05\times10^{-4}$ & $9.34\times10^{-4}$ & $2.62\times10^{-2}$ & $1.29\times10^{-2}$ &  &  &  & \tabularnewline
\cline{2-2} \cline{9-9} \cline{11-15} 
 & d.U. core &  &  &  &  &  &  & $1.16\times10^{-8}$ &  & $2.31\cdot10^{-5}$ & $\frac{5}{3}$ & $4$ & $1$ & $6$\tabularnewline
\hline 
\end{tabular}

\caption{\label{tab:parameter-values}Parameters characterizing the neutron
star considered in this work. The constants $\tilde{I}$, $\tilde{J}$,
$\tilde{S}$, $\tilde{V}$, $\tilde{C}_{V}$ and $\tilde{L}$ are
given in tab. \ref{tab:Integral-parameters.} using the generic normalization
scales $\Lambda_{{\rm QCD}}=1$ GeV and $\Lambda_{{\rm EW}}=100$
GeV and the temperature exponents $\sigma$, $\delta$, $\upsilon$
and $\theta$ are defined by eqs.~(\ref{eq:viscosity-parametrization})
and (\ref{eq:thermal-parametrization}).}
\end{table*}

\section{Pulsar Evolution equations}

The evolution equations are obtained from energy and angular momentum
conservation laws \cite{Owen:1998xg,Levin:1999ApJ...517..328L,Ho:1999fh}
and take the form

\begin{align}
\frac{d\alpha}{dt} & =-\alpha\left(\frac{1}{\tau_{G}}+\frac{1}{\tau_{V}}\left(\frac{1-Q\alpha^{2}}{1+Q\alpha^{2}}\right)\right)\:,\label{eq:alpha-equation}\\
\frac{d\Omega}{dt} & =-\frac{2\Omega Q\alpha^{2}}{\tau_{V}}\frac{1}{1+Q\alpha^{2}}\:,\label{eq:Omega-equation}\\
\frac{dT}{dt} & =-\frac{1}{C_{V}}\left(L_{\nu}-P_{V}\right)\:,\label{eq:T-equation}
\end{align}
with $Q\equiv3\tilde{J}/(2\tilde{I})$ and in terms of the viscous
dissipated power $P_{V}=P_{S}+P_{B}+\cdots$ and damping time $1/\tau_{V}\equiv1/\tau_{S}+1/\tau_{B}+\cdots$,
where the dots denote possible other dissipative mechanisms, like
boundary layer effects. Here we include the reheating due to the dissipated
power $P_{V}$ within the star, that was partly neglected in \cite{Owen:1998xg}.
Using the previous bounds on the parameters $\tilde{I}$ and $\tilde{J}$
we see that $Q<81/(112\pi)\approx0.23$ so that the factors $1\pm Q\alpha^{2}$
are only relevant for large amplitude modes with $\alpha\gtrsim1$,
i.e.\ in a regime where the perturbative approximation for the r-mode
\cite{Lindblom:1999yk} breaks down anyway. The r-mode is unstable
when the right hand side of the amplitude equation eq.~(\ref{eq:alpha-equation})
is positive, i.e.\ when $\left.\tau_{V}\right|_{\alpha=0}\geq-\tau_{G}$,
where the equality defines the boundary of the instability region
\cite{Alford:2010fd}. Since the r-mode will grow once it enters the
instability region the evolution requires a non-linear mechanism to
saturate the amplitude at a finite value. A simple possibility is
the suprathermal enhancement of the bulk viscosity \cite{Reisenegger:2003pd,Alford:2010gw,Alford:2011pi}.
Unfortunately when only the damping in the core is considered, this
mechanism saturates the mode only at rather large amplitudes. The
bulk viscosity contribution of the inner crust might change this since
the outer regions of the star are strongly weighted by the suprathermal
viscosity \cite{Alford:2011pi}, but the bulk viscosity has not been
computed at this point. Another promising mechanism is the non-linear
coupling of the r-mode to other daughter modes that are subsequently
damped by viscosity \cite{Arras:2002dw,Bondarescu:2008qx,Bondarescu:2013xwa}.
Further possibilities include non-linear hydrodynamic effects \cite{Lindblom:2000az,Kastaun:2011yd}
or turbulent crust-core boundary layer damping \cite{Wu:2000qy} but
at this point it is not settled which of these mechanisms will actually
saturate the mode%
\footnote{A reason for this is that because of the complexity of the system
all of these analyses have to make simplifications. For instance in
mode coupling scenarios the possible interactions have to be restricted
in the analysis, whereas in hydrodynamical simulations the radiation
reaction force needs to be artificially increased to make the computation
numerically feasible, see also \cite{Arras:2002dw}.%
}. Therefore, we follow here the approach pursued in \cite{Owen:1998xg}
and will not study a particular saturation mechanism but simply assume
that a corresponding mechanism operates and saturates the r-mode at
a particular amplitude $\alpha_{{\rm sat}}$. In contrast to the above
explicit mechanisms, $\alpha_{{\rm sat}}$ is in this case an unknown
parameter that may depend on $\Omega$ or $T$ and we will study the
dependence of our results on it below.

\begin{figure}
\includegraphics{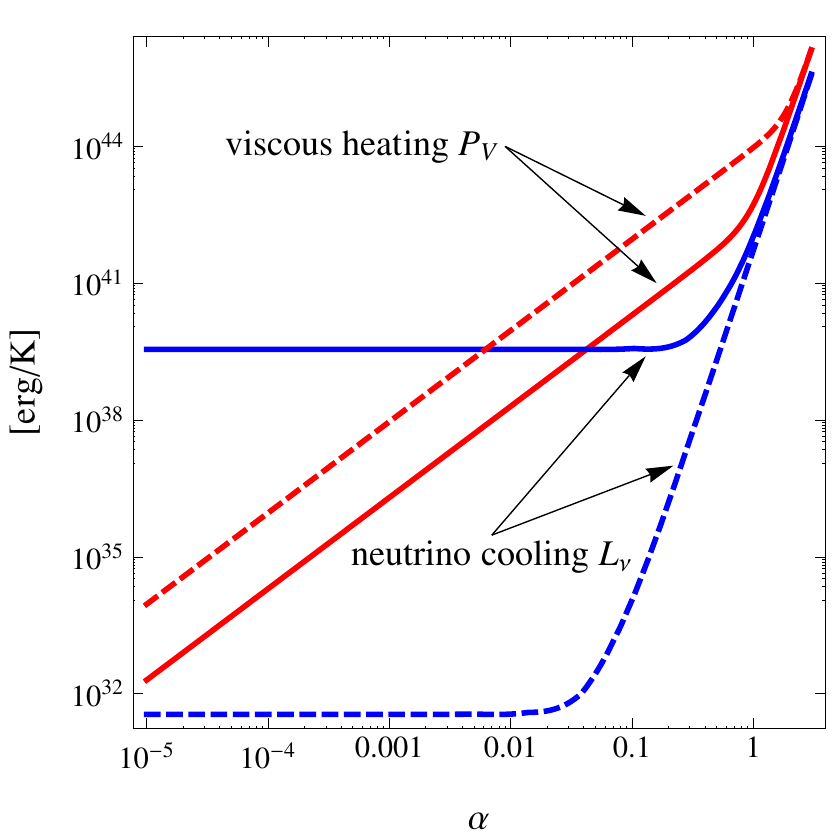}

\caption{\label{fig:heating-vs-cooling}The total neutrino luminosity compared
to the viscously dissipated power, as functions of the r-mode amplitude.
We show curves for a $1.4\, M_{\odot}$ APR neutron star spinning
at half its Kepler frequency and for the two temperatures $T=10^{8}\,{\rm K}$
(dashed) and $T=10^{9}\,{\rm K}$ (solid). At lower frequencies the
suprathermal regime is only reached at even larger amplitudes \cite{Alford:2011pi,Alford:2010gw}.}
\end{figure}

At saturation $d\alpha/dt=0$, so from eq.~(\ref{eq:alpha-equation}),

\begin{equation}
\frac{1}{\tau_{V}}=\frac{1}{\tau_{G}}\left(\frac{1+Q\alpha^{2}}{1-Q\alpha^{2}}\right)\xrightarrow[\alpha\ll1/\sqrt{Q}]{}\frac{1}{\tau_{G}}\:,\label{eq:saturation-condition}
\end{equation}
and likewise $P_{V}\to P_{G}.$ This simply says that at saturation
there must be a sufficiently strong source of dissipation to overcome
the gravitational instability. Making the replacements in eqs.~(\ref{eq:alpha-equation}),
(\ref{eq:Omega-equation}) and (\ref{eq:T-equation}) leaves the reduced
set

\begin{align}
\frac{d\Omega}{dt} & =-\frac{2\Omega}{\left|\tau_{G}\right|}\frac{Q\alpha_{{\rm sat}}^{2}}{1-Q\alpha_{{\rm sat}}^{2}}\approx-\frac{2Q\alpha_{{\rm sat}}^{2}\Omega}{\left|\tau_{G}\right|}\:,\label{eq:Omega-equation-saturation}\\
\frac{dT}{dt} & =-\frac{1}{C_{V}}\left(L_{\nu}\!+\! P_{G}\left(\frac{1\!+\! Q\alpha_{{\rm sat}}^{2}}{1\!-\! Q\alpha_{{\rm sat}}^{2}}\right)\right)\approx-\frac{1}{C_{V}}\left(L_{\nu}\!+\! P_{G}\right)\:,\label{eq:T-equation-saturation}
\end{align}
where, as argued before, the approximate expressions hold over nearly
the entire range of physically reasonable amplitudes. Note that compared
to \cite{Owen:1998xg} we take into account that according to eq.~(\ref{eq:saturation-condition})
the saturation mechanism dissipates a significant amount of energy
that heats the star. The only way that this could be avoided is if
some of the corresponding energy is directly, non-thermally radiated
away, e.g.\ in neutrinos or electromagnetic radiation. In all proposed
mechanisms, like the non-linear enhancement of the bulk viscosity
\cite{Reisenegger:2003pd,Alford:2011pi}, mode-coupling to viscously
damped daughter modes \cite{Arras:2002dw,Bondarescu:2008qx} or nonlinear
hydrodynamic effects \cite{Lindblom:2000az,Kastaun:2011yd}, such
a radiative component is negligible and nearly all the dissipated
energy eventually ends up as heat. Photons are only emitted from the
surface and the emission process is not affected in any significant
way by oscillation modes. For neutrinos the additional emission due
to a global mode is described by the amplitude-dependent, suprathermal
regime of the neutrino emissivity \cite{Reisenegger:1994be,Alford:2011df}.
The neutrino luminosity of a neutron star is compared to the viscously
dissipated power in fig.~\ref{fig:heating-vs-cooling}. As can be
seen the suprathermal neutrino luminosity, characterized by the steeply
rising part, becomes relevant compared to the shear viscosity only
at amplitudes $\alpha=O\!\left(1\right)$; whereas at lower amplitudes
the energy loss by non-thermal, induced neutrino emission is entirely
negligible. Daughter modes arising in mode-coupling mechanisms have
lower amplitudes and therefore nonlinear processes are even less likely
to play a role. In conclusion, the energy that is continuously fed
into the mode from the rotational energy reservoir must be completely
dissipated to keep the amplitude constant which continuously heats
the star.

The complete information on the saturation mechanism is then encoded
in the saturation amplitude which can be a general function $\alpha_{{\rm sat}}\!\left(T,\Omega\right)$.
This function has to vanish at the boundary of the instability region.
Due to the strong power law dependence of the corresponding time scales
$\tau_{G}$ and $\tau_{V}$ on $T$ and $\Omega$, it drops sharply
over a very narrow region, as has been explicitly shown in \cite{Alford:2011pi}.
For the spindown evolution this narrow boundary region is far less
important than the much larger interior of the instability region.
In the interior the amplitude will depend much more weakly on $T$
and $\Omega$ and we make the simple power law ansatz 
\begin{equation}
\alpha_{{\rm sat}}=\hat{\alpha}_{{\rm sat}}T^{\beta}\Omega^{\gamma}\:.\label{eq:alpha-ansatz}
\end{equation}
For instance the semi-analytic expression for the saturation amplitude
due to suprathermal bulk viscosity has such a power law dependence
on the frequency and proves to be even independent of temperature,
i.e.\ $\beta=0$ \cite{Alford:2011pi}. Since the only temperature
dependence in eq.~(\ref{eq:Omega-equation-saturation}) comes from
the implicit dependence via $\alpha_{sat}$, for $\beta=0$ the spindown
evolution is not affected at all by the thermal evolution. This applies
in particular to the model employed in \cite{Owen:1998xg} where the
saturation amplitude is assumed to be constant throughout the instability
region, i.e. independent of $T$ and $\Omega$. For a general saturation
mechanism the saturation amplitude could be a more complicated function,
but as in the analysis of the instability boundary \cite{Alford:2010fd}
the above form should hold at least locally in different regions in
the $T$-$\Omega$ space.

To analyze the above equations it is useful to introduce characteristic
evolution time scales

\begin{equation}
\tau_{\alpha}\equiv-\alpha\left(\!\frac{d\alpha}{dt}\!\right)^{-1},\;\tau_{\Omega}\equiv-\Omega\left(\!\frac{d\Omega}{dt}\!\right)^{-1},\;\tau_{T}\equiv-T\left(\!\frac{dT}{dt}\!\right)^{-1}.\label{eq:evolution-time-scales}
\end{equation}
A young compact star is born in a core bounce supernova with a large
temperature $T>10^{11}$ K. We will assume that the initial frequency
is above the minimum frequency of the instability region \cite{Alford:2010fd}.
Initially $\tau_{T}\sim T^{-\left(\theta-\upsilon-1\right)}$ is small
so that the star cools very fast due to neutrino emission and quickly
enters the instability region. There by definition $\left|\tau_{G}\right|<\left.\tau_{V}\right|_{\alpha=0}$
and (while the star cools further) the r-mode will grow with time
scale $\tau_{\alpha}\approx\tau_{G}$. For a compact star spinning
with a millisecond period this scale is of the order of seconds, so
that the amplitude quickly grows until it is saturated at some $\alpha_{{\rm sat}}$.
From this point on the fast evolution of $\alpha$ stops and it changes
slowly with temperature and frequency. Since cooling is initially
very quick but slows down considerably at lower temperatures, at some
point the temperature-independent dissipation eq.~(\ref{eq:saturation-condition})
required to saturate the mode will dominate. The important question
for the evolution is then if the spindown or the temperature change
is faster. To see this let us analyze the behavior when temperature
change due to dissipative heating (neglecting neutrino emission) $\tau_{T}^{\left(H\right)}$
is faster than the spindown, i.e.\ $\tau_{T}^{\left(H\right)}<\tau_{\Omega}$.
This condition yields a relation for the angular velocity of the star
which introducing the Kepler frequency $\Omega_{K}=\frac{4}{9}\sqrt{2\pi G\rho_{0}}$
can be written in the form

\begin{align}
\frac{\Omega}{\Omega_{K}}\! & >\!\frac{27}{8\pi}\sqrt{\frac{\tilde{C}_{V}}{\tilde{I}\left(1-Q\alpha_{{\rm sat}}^{2}\right)}\frac{R}{R_{S}}\frac{\Lambda_{{\rm QCD}}^{4}}{\bar{\rho}}\left(\frac{T}{\Lambda_{{\rm QCD}}}\right)^{\upsilon+1}}\nonumber \\
 & \xrightarrow[\upsilon=1]{}const\frac{T}{\Lambda_{QCD}}\:,\label{eq:thermal-spindown-boundary}
\end{align}
where in the second line we use the fact that for dense matter in
general the temperature exponent arising in the specific heat eq.~(\ref{eq:thermal-parametrization})
obeys $\upsilon\geq1$, and larger values make the right hand side
even smaller. For a compact star the constant is $O\!\left(1\right)$,
since the dimensionless parameters $\tilde{C}_{V}$ and $\tilde{I}$
are normalized to also be $O\!\left(1\right)$, the characteristic
scale for the average energy density $\bar{\rho}$ of strongly interacting
matter is $\Lambda_{{\rm QCD}}^{4}$, the radius of a neutron star
is close to its Schwarzschild radius $R_{S}=2GM$ and the root further
decreases deviations from unity. However, since a compact star is
a degenerate system, after a short time it reaches a temperature where
$T/\Lambda_{{\rm QCD}}\lesssim O\!\left(10^{-3}\right)$. Therefore,
taking into account the $O\left(1\right)$ prefactor, the viscous
heating should be faster than the spindown for frequencies $\Omega/\Omega_{K}\gtrsim10^{-2}$.
Since r-modes are only present within the instability region, this
should be compared to the minimum of the instability region $\Omega_{{\rm min}}$,
for which a semi-analytic expression is given in \cite{Alford:2010fd}.
There it was shown that this minimum is extremely insensitive to the
microscopic details which yields a general limit of roughly $\Omega_{{\rm min}}/\Omega_{K}\gtrsim1/20$.
Therefore, at r-mode saturation within the instability region the
dissipative heating is always faster than the spindown. Note that
this conclusion is generic and to leading order independent of both
the saturation amplitude and the composition of the star.

\section{Semi-analytic solution of the r-mode evolution\label{sec:Semi-analytic-expressions}}

\subsection{Endpoint of the evolution}

Since for a compact star the thermal evolution is faster than the
spindown, the evolution of the star should follow a curve $\Omega_{hc}\!\left(T\right)$
where it is always thermally in a steady state where neutrino cooling
equals dissipative heating $L_{\nu}+P_{G}=0$ \cite{Bondarescu:2008qx}.
With the semi-analytic expressions eqs.~(\ref{eq:gravitational-power})
and (\ref{eq:neutrino-luminosity}) this equation can be solved explicitly
and yields for the dominant $m=2$ r-mode

\begin{equation}
\Omega_{hc}\!\left(T\right)=\left(\frac{3^{8}5^{2}}{2^{15}}\frac{\tilde{L}\Lambda_{{\rm QCD}}^{9-\theta}T^{\theta-2\beta}}{\tilde{J}^{2}\Lambda_{{\rm EW}}^{4}GM^{2}R^{3}\hat{\alpha}_{{\rm sat}}^{2}}\right)^{1/\left(8+2\gamma\right)}\:.\label{eq:steady-state-curve}
\end{equation}
The evolution should therefore follow this curve until the star leaves
the instability region. As can be seen in fig.~\ref{fig:instability-evo},
the boundary of the instability region of a neutron star has different
segments: a low-temperature boundary outside of which shear viscosity
damps the mode and an intermediate-temperature boundary where bulk
viscosity is the dominant damping mechanism. Since the bulk viscosity
features a resonant behavior and decreases again at very large temperatures
$T\gtrsim10^{11}\,{\rm K}$, there a is third high-temperature boundary
above which the r-mode is unstable (see fig.~3 of \cite{Alford:2010fd}),
but for neutron stars this segment is quite likely physically irrelevant
since the stars cools below it quickly. Semi-analytic expressions
for these segments of the boundary of the instability region have
been derived in \cite{Alford:2010fd}. The form of the boundary depends
strongly on the form of matter \cite{Ho:2011tt,Haskell:2012}. When
exotic forms of matter are present then in the relevant temperature
regime there are so called stability windows, caused by the resonant
behavior of the bulk viscosity of the corresponding phase, where the
star is stable against developing an r-mode up to large frequencies
(see fig.~3 of \cite{Alford:2010fd}). In case the steady-state curve
formally lies in such a stability window it would clearly be irrelevant
for the spindown%
\footnote{As just noted, for standard neutron stars the corresponding stability
window lies at very large temperatures $T>10^{10}$ K where the neutrino
cooling still strongly dominates.%
}. Then the evolution can take a long time, since the strong heating
can push the star out of the instability region before the r-mode
reaches a large amplitude, and other spindown sources will likely
dominate. Therefore we will restrict ourselves here to the case of
pure neutron stars. The general evolution for different forms of matter
will be discussed elsewhere.

In the case of neutron stars the intermediate-temperature boundary
of the instability region has a steeper slope than the steady state
curve which thereby can only intersect the low-temperature boundary.
In the case of a neutron star the low-temperature boundary $\Omega_{lb}\!\left(T\right)$
is determined by the condition $\tau_{S}\!=\!\tau_{G}$ and the corresponding
semi-analytic solution is \cite{Alford:2010fd}

\begin{equation}
\Omega_{lb}\!\left(T\right)=\left(\frac{3^{8}5^{3}}{2^{17}\pi}\frac{\tilde{S}\Lambda_{{\rm QCD}}^{3+\sigma}}{\tilde{J}^{2}GM^{2}R^{3}T^{\sigma}}\right)^{1/6}\:.\label{eq:left-boundary}
\end{equation}
 Equating eqs.~(\ref{eq:steady-state-curve}) and (\ref{eq:left-boundary}),
the final angular velocity $\Omega_{f}$ where the evolution leaves
the instability region is given by

\begin{align}
 & \Omega_{f}=\Biggl(\left(\frac{3^{8}5^{3}}{2^{17}\pi}\right)^{\theta+\sigma-2\beta}\left(\frac{4\pi}{5}\right)^{\sigma}\Bigg.\label{eq:final-frequency}\\
 & \times\Bigg.\frac{\tilde{S}^{\theta-2\beta}\tilde{L}^{\sigma}\Lambda_{{\rm QCD}}^{3\theta+9\sigma-6\beta-2\beta\sigma}}{\left(\tilde{J}^{2}GM^{2}R^{3}\right)^{\theta+\sigma-2\beta}\Lambda_{{\rm EW}}^{4\sigma}\hat{\alpha}_{{\rm sat}}^{2\sigma}}\!\Biggr)^{\!1/\left(6\theta+8\sigma+2\gamma\sigma-12\beta\right)},\nonumber 
\end{align}
and the corresponding temperature $T_{f}$ reads

\begin{align}
 & T_{f}=\Biggl(\left(\frac{3^{8}5^{3}}{2^{17}\pi}\right)^{2+2\gamma}\left(\frac{5}{4\pi}\right)^{6}\Biggr.\label{eq:final-temperature}\\
 & \!\times\Biggl.\frac{\tilde{S}^{8+2\gamma}\Lambda_{{\rm EW}}^{24}\Lambda_{{\rm QCD}}^{-30+6\theta+8\sigma+6\gamma+2\gamma\sigma}\hat{\alpha}_{{\rm sat}}^{12}}{\tilde{L}^{6}\left(\tilde{J}^{2}GM^{2}R^{3}\right)^{2\left(1+\gamma\right)}}\!\Biggr)^{\!1/\left(6\theta+8\sigma+2\gamma\sigma-12\beta\right)}\!.\nonumber 
\end{align}
Note that, if we had used a different steady-state curve $\Omega_{hc}^{\left(S\right)}$
where the heating comes only from shear viscosity, as assumed in \cite{Owen:1998xg},
or if we had included the bulk viscosity as well, the rest of the
evolution could be very different, but we would have obtained exactly
the same result for the endpoint eqs.~(\ref{eq:final-frequency})
and (\ref{eq:final-temperature}) of the evolution . This surprising
feature comes about since the shear viscosity is strongly dominant
compared to the bulk viscosity on the left boundary and this boundary
is precisely defined by the condition $\tau_{S}\!=\!\tau_{G}$ which
implies that dissipated power equals the power radiated in gravitational
waves $P_{S}\!=\! P_{G}$. In this sense the result eq.~(\ref{eq:final-frequency})
is universal and does not depend on assumptions about the dissipative
aspects of the saturation mechanism.

The above expressions involve the star constants given in tab.~\ref{tab:parameter-values}.
In order to give these expressions in a more physical form we write
them relative to a fiducial reference model chosen as a $1.4\, M_{\odot}$
neutron star with an APR equation of state \cite{Akmal:1998cf} given
in the first row of tab. \ref{tab:parameter-values} and denoted by
a subscript ``${\rm fid}$''. In case of a constant, $T$- and
$\Omega$-independent, saturation amplitude \cite{Owen:1998xg} $\beta\!=\!\gamma\!=\!0$
this gives for a \emph{standard} neutron star $(NS)$ with shear viscosity
due to leptonic processes and neutrino emission due to modified Urca
reactions

\begin{align}
 & T_{f}^{\left(NS\right)}\approx\left(1.26\cdot10^{9}\,\mathrm{K}\right)\left(\!\frac{\tilde{S}}{\tilde{S}_{{\rm fid}}}\!\right)^{\!3/23}\!\left(\!\frac{\tilde{L}}{\tilde{L}_{{\rm fid}}}\!\right)^{\!-9/92}\!\left(\!\frac{\tilde{J}}{\tilde{J}_{{\rm fid}}}\!\right)^{\!-3/46}\nonumber \\
 & \quad\times\left(\!\frac{M}{1.4\, M_{\odot}}\!\right)^{-3/46}\!\left(\!\frac{R}{11.5\,\mathrm{km}}\!\right)^{-9/92}\!\alpha_{{\rm sat}}^{9/46}\,,\label{eq:final-temperature-NS}\\
 & f_{f}^{\left(NS\right)}\approx\left(61.4\,\mathrm{Hz}\right)\left(\!\frac{\tilde{S}}{\tilde{S}_{{\rm fid}}}\!\right)^{\!3/23}\!\left(\!\frac{\tilde{L}}{\tilde{L}_{{\rm fid}}}\!\right)^{\!5/184}\!\left(\!\frac{\tilde{J}}{\tilde{J}_{{\rm fid}}}\!\right)^{\!-29/92}\nonumber \\
 & \quad\times\left(\!\frac{M}{1.4\, M_{\odot}}\!\right)^{-29/92}\!\left(\!\frac{R}{11.5\,\mathrm{km}}\!\right)^{-87/184}\!\alpha_{{\rm sat}}^{-5/92}\,.\label{eq:final-frequency-NS}
\end{align}
First it is interesting to note that all these expressions depend
neither on the specific heat nor directly on the bulk viscosity of
the considered form of matter. The latter is at first sight surprising
since the bulk viscosity is the dominant dissipation mechanism at
high temperatures and low amplitudes. But during the spindown evolution
the dissipation is dominated by the appropriate saturation mechanism
while the amount of dissipation is according to eqs.~(\ref{eq:time-scales})
and (\ref{eq:saturation-condition}) determined entirely by gravitational
physics. This is particularly favorable since it eliminates the complications
arising from the fact that the damping due to bulk viscosity requires
the density fluctuation of the r-mode eq.~(\ref{eq:delta-Sigma})
which is only non-vanishing to second order in $\Omega$ \cite{Lindblom:1999yk}
and involves significant uncertainties. Even more striking is the
extreme insensitivity of these result to the remaining microphysics
included in the constants $\tilde{S}$ and $\tilde{L}$. The neutrino
emissivity constant $\tilde{L}$, in particular, depends on poorly
known strong interaction corrections to the weak processes \cite{Friman:1978zq}.
Yet, the remarkably small power $5/184$ arising in the important
expression for the final angular velocity $\Omega_{f}^{\left(NS\right)}$
almost eliminates this uncertainty. The dependence of $\Omega_{f}^{\left(NS\right)}$
on the shear viscosity constant $\tilde{S}$ is somewhat stronger,
but $\tilde{S}$ is dominated by electromagnetic leptonic processes
\cite{Shternin:2008es} that are well under theoretical control%
\footnote{There is a hadronic contribution to the shear viscosity as well but
the latter is strongly subleading at temperatures relevant to young
neutron stars due to the different power law dependence.%
} and depends on the equation of state only via the lepton densities.
The dependence on the parameters $\tilde{J}$, $M$ and $R$ which
encode macroscopic properties of the star is stronger, but they vary
only within narrow margins. This insensitivity to the underlying parameters
is similar to the case of the minimum of the instability region \cite{Lindblom:1998wf}
analyzed previously in \cite{Alford:2010fd}, yet it is fortunately
even more pronounced in the physically important case eq.~(\ref{eq:final-frequency-NS}).
The most striking feature of these expressions, however, is that the
insensitivity extends even to the dependence on $\alpha_{{\rm sat}}$,
which is theoretically uncertain by many orders of magnitude \cite{Alford:2011pi,Bondarescu:2008qx,Lindblom:2000az,Kastaun:2011yd}
and therefore represents the main uncertainty in the analysis. Here
the similarly small power $5/92$ dramatically reduces this dependence
and thereby allows a quantitative comparison with observational data
below.

To get an idea of the uncertainty in $f_{f}^{\left(NS\right)}$ we
estimate the error ranges of the individual parameters by symmetric
logarithmic error bands - i.e.\ we allow for the corresponding parameter
to be smaller or larger than the APR reference star by a given factor.
The shear viscosity parameter $\tilde{S}$ due to leptonic processes
\cite{Shternin:2008es} is uncertain within a factor of two of $\tilde{S}_{{\rm fid}}$
and the neutrino emissivity parameter $\tilde{L}$ due to modified
Urca processes \cite{Friman:1978zq} within an order of magnitude
of $\tilde{L}_{{\rm fid}}$. For $\tilde{J}$ we have rigorous limits
eq.~(\ref{eq:J-tilde-bounds}). An upper limit for the mass of a
neutron star is $\lesssim2.5\, M_{\odot}$ and an upper limit for
the radius of a star rotating with millisecond periods about $\lesssim15$
km. Since the upper bound for $\tilde{J}$ is for a constant density
profile which is far from the situation in a neutron star and, due
to cancellations, the dependence on the mass and radius turns out
to be significantly weaker than these individual individual errors
suggest, see table \ref{tab:results} below, we consider half the
above ranges for the individual logarithmic uncertainties for $\tilde{J}$,
$M$ and $R$ as a more realistic estimate. Under the assumption that
these errors are independent we find

\begin{equation}
f_{f}^{\left(NS\right)}\approx\left(61.4\pm9.4\right)\mathrm{Hz}\,\,\alpha_{{\rm sat}}^{-5/92}\:.\label{eq:frequency-uncertainty}
\end{equation}
We want to stress that this is just an estimate of the likely error
range and not a statistical measure of well defined significance.

Different classes of compact stars, containing other phases of dense
matter with different low energy degrees of freedom, feature different
values for the exponents in eqs.~(\ref{eq:viscosity-parametrization})
and (\ref{eq:thermal-parametrization}) and can lead to qualitatively
different behavior. An example is a heavy star where direct Urca reactions
are kinematically allowed in an inner core and the parameters change
according to table \ref{tab:parameter-values} which increases the
final frequency beyond the above error range. When exotic forms of
matter are present, like hyperon or quark matter, the enhanced dissipation
of these phases leads to stability windows \cite{Jaikumar:2008kh,Alford:2010fd,Ho:2011tt,Haskell:2012}.
In this case the boundary of the stability window in the relevant
temperature regime is determined by the bulk viscosity of the exotic
phase, so that eqs.~(\ref{eq:final-frequency}) and (\ref{eq:final-temperature})
are not valid. The steady state curve then ends at a higher frequency
due to the instability window where the star leaves the unstable regime
and cools until it reaches the lower boundary of the stability window.
Since at these lower temperatures r-mode heating dominates neutrino
cooling the star is repeatedly pushed out of the instability region
once its amplitude becomes large and the residual spindown takes much
longer \cite{Madsen:1999ci,Andersson:2001ev}. The important point
is however, that the final frequency which is in this case given by
the minimum of the instability region and for which a similarly precise
semi-analytic result exists \cite{Alford:2010fd} is systematically
larger for exotic forms of matter since the strong bulk viscosity
of the exotic phase dominates the shear viscosity down to lower temperatures.
Therefore, the above expressions eqs.~(\ref{eq:final-frequency-NS})
and (\ref{eq:frequency-uncertainty}) provide lower bounds for the
frequency to which an arbitrary compact star can be spun down due
to the r-mode instability.

\subsection{Dynamical evolution and time scales}

Next let us actually solve the evolution equations to obtain the corresponding
evolution time scales. The initial cooling time in the stable regime
before the evolution enters the instability region $t_{sc}$ results
from solving the thermal equation eq.~(\ref{eq:T-equation}), in
the absence of viscous heating, down to the temperature determined
by the semi-analytic expression for the intermediate-temperature boundary
of the instability region given in \cite{Alford:2010fd}. The result
reads

\begin{align}
t_{sc} & =\frac{1}{\theta-\upsilon-1}\left(\frac{3^{3}5^{2}}{2^{12}7\pi\kappa^{2}}\right)^{\left(\theta-\upsilon-1\right)/\delta}\label{eq:initial-time}\\
 & \quad\times\left(\frac{\tilde{V}_{m}^{\theta-\upsilon-1}\tilde{C}^{\delta}\Lambda_{{\rm QCD}}^{9\left(\theta-\upsilon-1\right)-5\delta}R^{\theta-\upsilon-1}}{\tilde{L}^{\delta}\Lambda_{{\rm EW}}^{4\left(\theta-\delta-\upsilon-1\right)}\left(\tilde{J}^{2}GM^{2}\Omega_{i}^{4}\right)^{\theta-\upsilon-1}}\right)^{1/\delta}\nonumber 
\end{align}
in terms of the initial angular velocity $\Omega_{i}$ and depends
both on the bulk viscosity and the specific heat. For a standard neutron
star with shear viscosity due to leptonic processes and neutrino emission
due to modified Urca reactions this gives

\begin{align}
t_{sc}^{\left(NS\right)} & \approx\left(7.83\cdot10^{-2}\,\mathrm{s}\right)\left(\!\frac{\tilde{V}}{\tilde{V}_{{\rm fid}}}\!\right)\left(\!\frac{\tilde{J}}{\tilde{J}_{{\rm fid}}}\!\right)^{\!-2}\!\left(\!\frac{\tilde{C}}{\tilde{C}_{{\rm fid}}}\!\right)\left(\!\frac{\tilde{L}}{\tilde{L}_{{\rm fid}}}\!\right)^{\!-1}\nonumber \\
 & \qquad\times\left(\!\frac{M}{1.4\, M_{\odot}}\!\right)^{\!-2}\!\left(\!\frac{R}{11.5\,\mathrm{km}}\!\right)\!\left(\frac{f_{i}}{\mathrm{kHz}}\right)^{\!-4}\:.\label{eq:initial-time-NS}
\end{align}
This time is so short compared to the subsequent r-mode evolution
that it can safely be neglected. 

The spindown time $t_{sd}$ is obtained by solving the frequency equation
(\ref{eq:Omega-equation-saturation}) in the saturated regime. Even
though for r-mode emission the spindown evolution is coupled to the
thermal evolution when the saturation amplitude is temperature-dependent
the knowledge of the path of the evolution within the instability
region, determined by the analytic steady state solution eq.~(\ref{eq:steady-state-curve})
of the thermal equation, allows us to derive the effective spindown
equation. Inserting (the inverse of) the steady state relation eq.~(\ref{eq:steady-state-curve})
into the spindown equation (\ref{eq:Omega-equation-saturation}) we
find

\begin{align}
\frac{d\Omega}{dt}= & -\frac{2^{17}\pi}{3^{7}5^{2}}\frac{\tilde{J}^{2}GMR^{4}}{\tilde{I}}\left(\frac{2^{15}}{3^{8}5^{2}}\frac{\tilde{J}^{2}G\Lambda_{{\rm EW}}^{4}M^{2}R^{3}}{\tilde{L}\Lambda_{{\rm QCD}}^{9-\theta}}\right)^{\frac{2\beta}{\theta-2\beta}}\nonumber \\
 & \times\hat{\alpha}_{{\rm sat}}^{\frac{2\theta}{\theta-2\beta}}\Omega^{n_{rm}}\label{eq:effective-spindown-equation}
\end{align}
I.e. despite the coupling with the thermal evolution, the spindown
equation takes a standard power law form \cite{1983bhwd.book.....S},
yet with an \emph{effective} braking index

\begin{equation}
n_{rm}=\frac{\left(7\!+\!2\gamma\right)\theta\!+\!2\beta}{\theta\!-\!2\beta}=7\left(\!\frac{1\!+\!2\gamma/7\!+\!2\beta/\left(7\theta\right)}{1\!-\!2\beta/\theta}\!\right)\label{eq:effective-braking-index}
\end{equation}
The correction factor in parentheses changes the braking index from
the canonical value which is 7 for r-mode emission. The correction
factor takes the value $1$ for a constant saturation amplitude ($\beta\!=\!\gamma\!=\!0$),
but for general $\beta$ and $\gamma$ it deviates from $1$ and lowers
the braking index, since for proposed saturation mechanisms \cite{Arras:2002dw,Bondarescu:2008qx,Bondarescu:2013xwa,Alford:2011pi}
$\beta,\gamma<0$. Using these values we find that the correction
factor can decrease the braking index to nearly half of its canonical
value and therefore has a significant impact on the spindown evolution.
E.g. for mode coupling with damping due to shear viscosity \cite{Bondarescu:2013xwa},
where $\beta=-4/3$ and $\gamma=-2/3$, it reduces to $n_{rm}^{\left(mc\right)}=4$.
This is rather close to the canonical value of $3$ for electromagnetic
dipole radiation. Therefore, it should be harder than expected to
distinguish these different spindown scenarios based on their braking
indices. Actually, an even lower value is obtained for the recently
proposed scenario that the cutting of superconducting flux tubes by
superfluid vortices can saturate r-modes \cite{Haskell:2013hja},
where $\beta=0$, $\gamma\!=-3$. In this case the braking index takes
the extreme value $n_{rm}^{\left(fc\right)}=1$, which means that
the spindown is not even power-like but logarithmic and the general
solution given below does not a apply. However, the spindown evolution
decouples in this case and can be easily solved. The case of the suprathermal
bulk viscosity \cite{Alford:2011pi} is more complicated and will
be studied elsewhere, since the large amplitude enhancement of the
neutrino emissivity would here be relevant to determine the effective
braking index. As discussed before such effects are not relevant unless
suprathermal damping saturates the mode, which is not the case for
standard bulk viscosity from a neutron stars core.

Even though the spindown evolution is coupled to the thermal evolution
when the saturation amplitude is temperature-dependent, the knowledge
of the path of the evolution within the instability region at saturation
eq.~(\ref{eq:steady-state-curve}) allows us to solve the spindown
equation (\ref{eq:effective-spindown-equation}) fully analytically
for the power-law parameterization eq.~(\ref{eq:alpha-ansatz}) which
gives%
\footnote{As noted before, in general the saturation amplitude might only locally
have the power law from eq.~(\ref{eq:alpha-ansatz}). In this case
there are different regions in $T$-$\Omega$ space where the evolution
is locally described by eq.~(\ref{eq:spindown-solution}) and one
would in principle have to solve the evolution stepwise with appropriate
initial conditions. Since the spindown strongly slows down at late
times and becomes then effectively independent of the initial conditions
(the first term in the parenthesis of eq.~(\ref{eq:spindown-solution})
can be neglected), at late times eq.~(\ref{eq:spindown-solution})
is a good approximation even in the general case. It is sufficient
at late times to consider the power law dependence of the saturation
amplitude realized in the final evolution segment and it is irrelevant
that the dependence was different early in the evolution.%
}

\begin{align}
 & \Omega\!\left(t\right)=\left(\Omega_{i}^{-\frac{2\left(3+\gamma\right)\theta+4\beta}{\theta-2\beta}}+\frac{2^{18}\pi}{3^{7}5^{2}}\frac{\left(3+\gamma\right)\theta+2\beta}{\theta-2\beta}\frac{\tilde{J}^{2}GMR^{4}}{\tilde{I}}\right.\nonumber \\
 & \left.\times\!\left(\!\frac{2^{15}}{3^{8}5^{2}}\frac{\tilde{J}^{2}G\Lambda_{{\rm EW}}^{4}M^{2}R^{3}}{\tilde{L}\Lambda_{{\rm QCD}}^{9-\theta}}\!\right)^{\frac{2\beta}{\theta-2\beta}}\negmedspace\hat{\alpha}_{{\rm sat}}^{\frac{2\theta}{\theta-2\beta}}\left(t-t_{i}\right)\!\right)^{\!-\frac{\theta-2\beta}{2\left(3+\gamma\right)\theta+4\beta}}\negthickspace.\label{eq:spindown-solution}
\end{align}
An interesting aspect of this expression is that for a frequency-independent
saturation amplitude it does not depend on the cooling behavior which
reflects the previous observation that the thermal and spindown evolutions
decouple in this case. Eq.~(\ref{eq:spindown-solution}) has two
limits, at early times where the first term in the parenthesis dominates
and at late times where the second dominates. I.e. initially the amplitude
hardly changes whereas at late times the evolution becomes independent
of the initial conditions. The semi-analytic temperature evolution
at saturation is then obtained by inserting the frequency solution
eq.~(\ref{eq:spindown-solution}) in the inverse of the steady state
curve eq.~(\ref{eq:steady-state-curve}). In the late time limit
we find

\begin{align}
 & T\!\left(t\right)\approx\left(\left(\frac{3^{7}5^{2}}{2^{18}\pi}\frac{\theta-2\beta}{\left(3+\gamma\right)\theta+2\beta}\frac{\tilde{I}}{\tilde{J}^{2}GMR^{4}t}\right)^{8+2\gamma}\right.\nonumber \\
 & \left.\times\left(\frac{2^{15}}{3^{8}5^{2}}\frac{\tilde{J}^{2}G\Lambda_{{\rm EW}}^{4}M^{2}R^{3}}{\tilde{L}\Lambda_{{\rm QCD}}^{9-\theta}}\right)^{2\left(3+\gamma\right)}\hat{\alpha}_{{\rm sat}}^{-4}\right)^{\frac{1}{2\left(3+\gamma\right)\theta+4\beta}}\negmedspace.\label{eq:temperature-evolution}
\end{align}
For a standard neutron star in particular we find for the constant
saturation model in the late time limit $\Omega\!\left(t\right)\sim T\!\left(t\right)\sim t^{-1/6}$.

For the spindown time we find then in terms of the gravitational
time scale eq.~(\ref{eq:gravitational-power}) evaluated at the final
frequency

\begin{align}
t_{sd} & =-\frac{\theta-2\beta}{4\left(3+\gamma\right)\theta+8\beta}\frac{\tau_{G}\left(\Omega_{f}\right)}{Q\alpha_{{\rm sat}}^{2}\left(\Omega_{f}\right)}\left(\!1\!-\!\left(\frac{\Omega_{f}}{\Omega_{i}}\right)^{\!\frac{2\left(3+\gamma\right)\theta+4\beta}{\theta-2\beta}}\!\right)\nonumber \\
 & \xrightarrow[\Omega_{f}\ll\Omega_{i}]{}-\frac{{\cal C}}{6}\frac{\Omega_{f}}{\dot{\Omega}_{f}}\:.\label{eq:spindown-time}
\end{align}
where the dependence on the saturation model enters only via the constant
prefactor

\begin{equation}
{\cal C}\equiv\frac{1-2\beta/\theta}{1+\gamma/3+2\beta/\left(3\theta\right)}{\color{red}=\frac{6}{n_{rm}-1}}\label{eq:saturation-model-factor}
\end{equation}
which takes the value $1$ for a constant saturation amplitude ($\beta\!=\!\gamma\!=\!0$).
For general $\beta$ and $\gamma$ it deviates from $1$, but for
proposed saturation mechanisms \cite{Arras:2002dw,Bondarescu:2008qx,Bondarescu:2013xwa,Alford:2011pi}
$\beta,\gamma<0$ so that up to small corrections in $\Omega_{f}/\Omega_{i}$
this simply increases the spindown time compared to a constant saturation
amplitude by a constant factor. Since in hadronic matter $\theta\!=\!8$
these corrections are modest%
\footnote{This holds in general degenerate fermionic phases where $\theta\geq6$.%
} and for the exponents predicted by proposed saturation mechanisms
\cite{Arras:2002dw,Bondarescu:2008qx,Bondarescu:2013xwa,Alford:2011pi}
it amounts at most to a factor two. E.g.~in case of dissipation due
to suprathermal bulk viscosity \cite{Alford:2011pi} the spindown
time is longer compared to the saturation model with a constant amplitude
by a factor of $9/5$. Taking into account our ignorance of the complicated
r-mode saturation physics this simple and moderate dependence is most
welcome.

The above expression eq.~(\ref{eq:spindown-time}) looks similar
to the relation between the spindown time for other spindown mechanisms,
like magnetic dipole radiation or gravitational wave emission due
to deformations, and the characteristic pulsar time scale $-\Omega_{0}/\dot{\Omega}_{0}$
\cite{1983bhwd.book.....S}. Yet, the important point to note here
is, that $\Omega_{f}$ and $\dot{\Omega}_{f}$ are the values before
the star leaves the instability region which can be very different
from the observed values $\Omega_{0}$ and $\dot{\Omega}_{0}$ if
the observation is performed after the star has left the instability
region. The characteristic feature of the r-mode mechanism, namely
that it operates only for a finite time interval, renders the r-mode
spindown time completely independent of the characteristic pulsar
time scale and allows the former to be orders of magnitude smaller,
since the spindown rate can strongly drop from a large r-mode rate
to a much lower rate of other spindown mechanisms at the boundary
of the instability region.

Inserting the expression for the final frequency eq.~(\ref{eq:final-frequency-NS})
into eq.~(\ref{eq:spindown-time}) this yields for the constant saturation
model in case of a standard neutron star and for $\Omega_{f}\ll\Omega_{i}$

\begin{align}
 & t_{sd}^{\left(NS\right)}\approx\left(12.3\mathrm{\, y}\right)\left(\!\frac{\tilde{S}}{\tilde{S}_{{\rm fid}}}\!\right)^{\!-18/23}\!\!\left(\!\frac{\tilde{L}}{\tilde{L}_{{\rm fid}}}\!\right)^{\!-15/92}\!\!\left(\!\frac{\tilde{J}}{\tilde{J}_{{\rm fid}}}\!\right)^{\!-5/46}\nonumber \\
 & \times\left(\!\frac{\tilde{I}}{\tilde{I}_{{\rm fid}}}\!\right)\!\left(\!\frac{M}{1.4\, M_{\odot}}\!\right)^{\!41/46}\!\!\left(\!\frac{R}{11.5\,\mathrm{km}}\!\right)^{\!-107/92}\!\alpha_{{\rm sat}}^{-77/46}\:.\label{eq:spindown-time-NS}
\end{align}
Here the dependence on the parameters is stronger than for the endpoint
of the evolution eqs.~(\ref{eq:final-frequency}) and (\ref{eq:final-temperature}),
so that the duration is considerably more uncertain. In particular
it is very sensitive to the saturation amplitude $\alpha_{{\rm sat}}$.
If $\alpha_{{\rm sat}}$ becomes too small this time scale is so long
that other spindown mechanisms will dominate, as is clear from the
sizable observed spindown rate of young pulsars \cite{Manchester:2004bp}.
Employing the bounds eqs.~(\ref{eq:I-tilde-bounds}) and (\ref{eq:J-tilde-bounds})
as well as the full uncertainty ranges for the other underlying parameters
discussed before to estimate the uncertainty in the final frequency
eq.~(\ref{eq:frequency-uncertainty}) we find that the spindown time
could be smaller or larger by nearly an order of magnitude.

\section{Spindown of young pulsars\label{sec:Spindown-of-young}}

\subsection{Comparison of analytic and numeric results}

\begin{figure}
\includegraphics[scale=0.85]{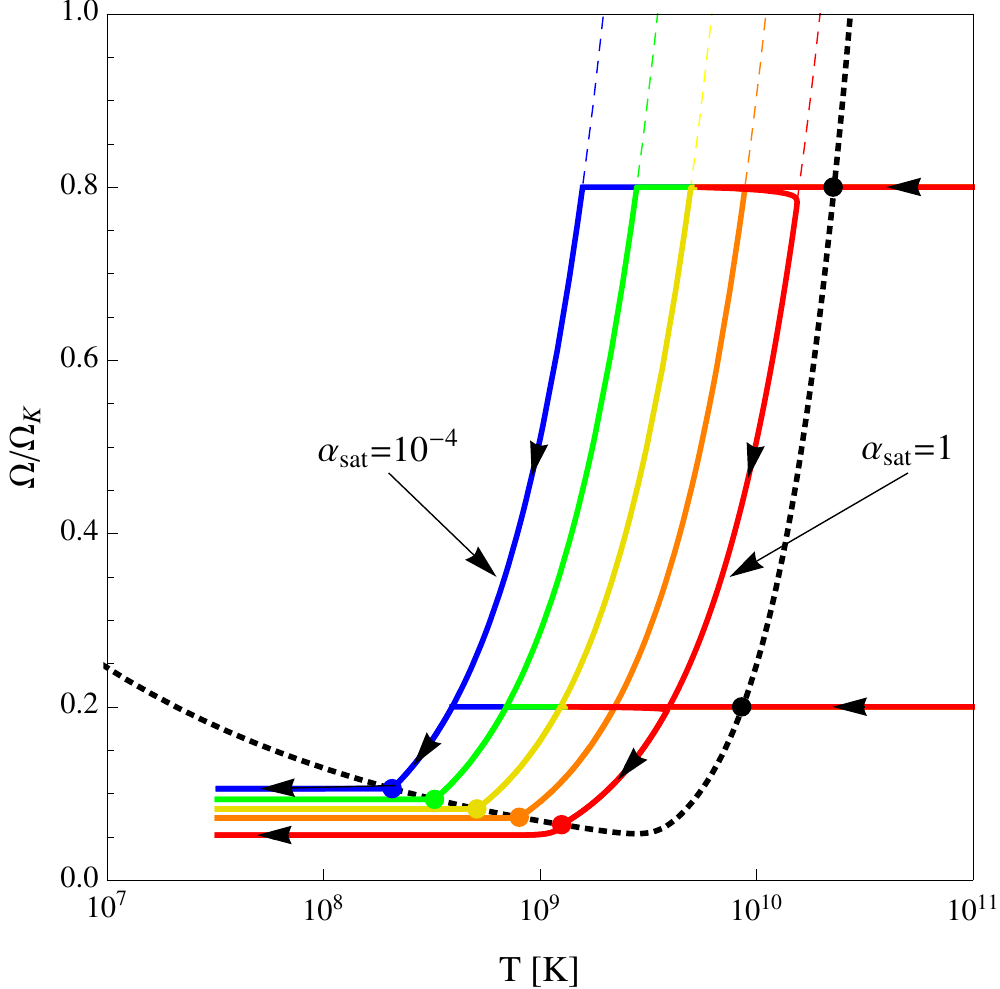}

\caption{\label{fig:instability-evo}The spindown evolution of a young $1.4\, M_{\odot}$
neutron star with an APR equation of state \cite{Akmal:1998cf} in
temperature-angular velocity space. The boundary of the instability
region of the fundamental $m=2$ r-mode is shown by the dotted curve.
The dashed curves (which are mostly hidden underneath the solid curves)
represent the steady state eq.~(\ref{eq:steady-state-curve}) where
heating equals cooling and are given for different r-mode amplitudes
ranging from $\alpha_{{\rm sat}}=10^{-4}$ (left) to $\alpha_{{\rm sat}}=1$
(right). The solid lines show the numerical solution of the evolution
equations for two fiducial initial spin frequencies $\Omega=0.8\,\Omega_{K}$
and $\Omega=0.2\,\Omega_{K}$. As can be seen, after an initial cooling
phase the evolution simply merges on the appropriate steady state
curve and follows it to the edge of the instability region. The dots
denote the semi-analytic results for the points of the spindown evolution
where the star enters and leaves the instability region, eqs.~(\ref{eq:final-temperature-NS})
and (\ref{eq:final-frequency-NS}).\textcolor{red}{}}
\end{figure}
The results for the evolution of the $1.4\, M_{\odot}$ neutron star
in temperature-angular velocity space is shown in fig.~\ref{fig:instability-evo}
for different values of the initial rotation angular velocity and
the saturation amplitude. As can be seen the numerical solution given
by the solid curves perfectly confirms the above picture. The star
initially cools until it reaches the instability region, where the
r-mode amplitude starts to grow. Once the amplitude is significant
the star starts spinning down but since the spindown time $\tau_{\Omega}$
is much longer than the cooling time $\tau_{T}$, the star cools further
until it reaches the steady state curve%
\footnote{For the largest saturation amplitudes the evolution reaches the steady
state temperature before its amplitude has saturated. Therefore, as
can be seen from the $\alpha_{{\rm sat}}\!=\!1$ curve in fig.~\ref{fig:instability-evo},
the star ``undercools'' to lower temperatures and then reheats back
to the steady state curve once the saturation amplitude is reached.%
} where the fast temperature evolution is stopped and the star evolves
along the steady state curve on the longer time scale $\tau_{\Omega}$.
Consequently, the evolution is completely independent of the initial
amplitude $\alpha_{{\rm min}}$ required to start the growth in eq.~(\ref{eq:alpha-equation}).
Such initial perturbations will clearly be present immediately after
the creation of the neutron star following a core bounce. As anticipated,
the numerical solution also proves that the evolution is to very good
accuracy independent of the initial frequency unless it is very close
to the final frequency. Therefore, similar to the thermal evolution,
the spindown evolution completely loses its memory of the initial
conditions at sufficiently late times. The saturation amplitude in
contrast provides the dominant uncertainty for the evolution and the
trajectory is shown for a range of possible saturation amplitudes.
Larger values lead to stronger dissipative reheating and therefore
spin down the star at larger temperatures so that the evolution leaves
the instability region at lower frequencies closer to the minimum
of the instability region. The dots are obtained from the analytic
expressions eqs.~(\ref{eq:final-temperature-NS}) and (\ref{eq:final-frequency-NS})
which agree with the numerical solution. Note again that we neglect
other spindown mechanisms like magnetic dipole radiation in this work
which would speed up the evolution and continue the magnetic spindown
after the star leaves the r-mode instability region.

In \cite{Owen:1998xg} the evolution was artificially stopped at a
temperature of $10^{9}$ K since superfluidity and non-perfect fluid
effects were expected to invalidate the analysis below. We do not
follow this ad hoc prescription here since it is not likely that these
effects have such a big influence on the spindown evolution for the
following reasons: One important effect of superfluidity is to affect
beta equilibration. Around the critical temperature it is enhanced
due to pair breaking, but at temperatures far below and small oscillation
amplitudes beta equilibration processes are strongly suppressed. Since
the superfluid gap depends strongly on density the entire star may
not be gapped at such temperatures and in this case in part of the
star the weak interactions would hardly be affected and the corresponding
parameters $\tilde{L}$ and $\tilde{V}$ would only moderately be
reduced. Even more, as has recently been shown in \cite{Alford:2011df},
superfluidity does not suppress weak processes at all at sufficiently
large amplitudes that are within the range we study here. Superfluidity
can also lead to an additional source of dissipation via mutual friction.
In contrast to the earlier generic analysis referred to in \cite{Owen:1998xg},
the later dedicated r-mode analysis \cite{Lindblom:1999wi} showed
that it is unlikely that mutual friction has such a dramatic impact
on the r-mode instability. In \cite{Haskell:2009fz} this effect was
studied, taking into account the large uncertainties in pairing and
the microscopic drag parameter. It was found that depending on these
uncertainties the effect can range from the unlikely extreme that
the r-mode is completely suppressed below a certain temperature to
a nearly negligible impact on the instability region compared to the
ungapped case. Therefore, it is interesting to study the present case
of standard dissipation sources in detail, since it sets the limit
on the frequency to which r-modes can spin down a star as well as
on the emitted gravitational wave signal, as discussed below. Any
additional dissipation source will reduce the r-mode instability and
correspondingly lead to larger final spin frequencies and a reduced
gravitational wave emission.

\begin{figure}
\includegraphics[scale=0.85]{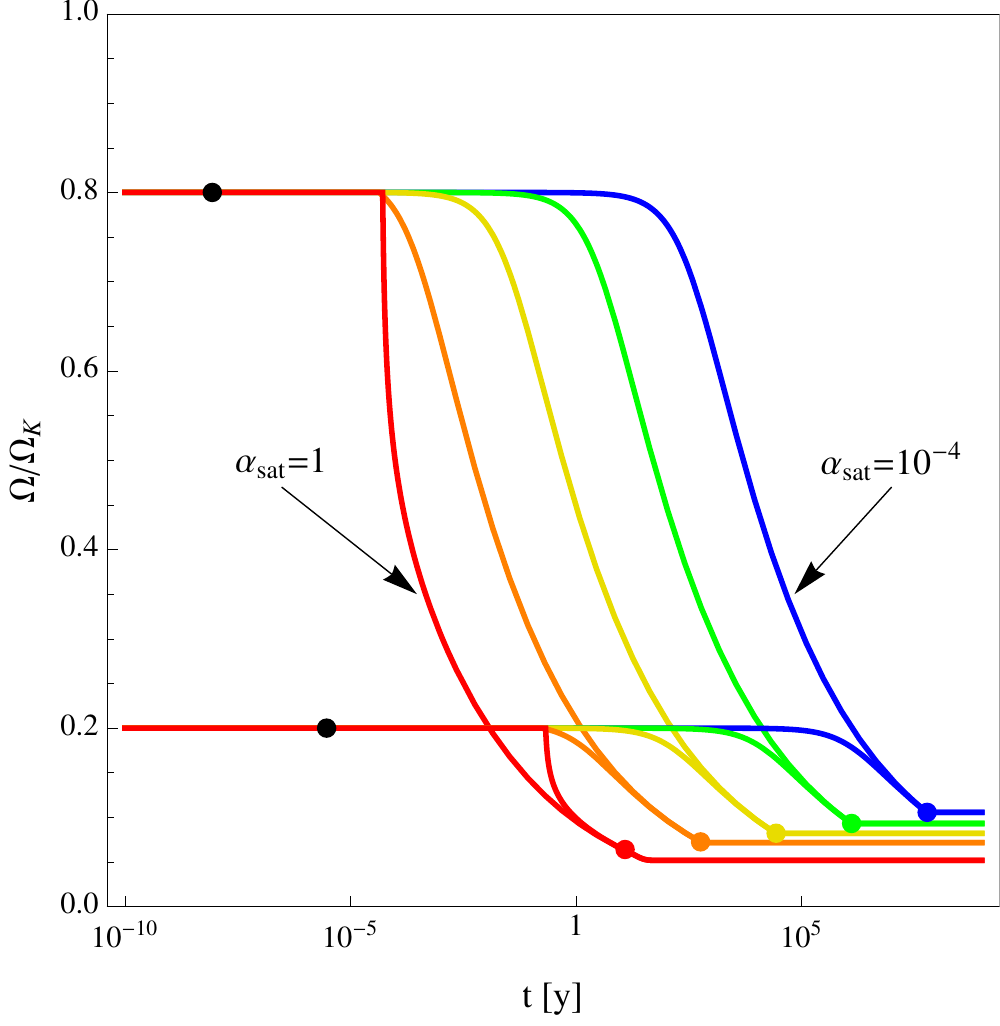}

\caption{\label{fig:frequency-evo}The spindown evolution of a young neutron
star. The solid lines show the evolution of a $1.4\, M_{\odot}$ star
with an APR equation of state and for different constant saturation
amplitudes ranging from $\alpha_{{\rm sat}}=1$ (left) to $\alpha_{{\rm sat}}=10^{-4}$
(right). The dots denote the analytic results eq.~(\ref{eq:initial-time-NS})
respectively eqs.~(\ref{eq:spindown-time-NS}) and (\ref{eq:final-frequency-NS})
for the point of the spindown evolution where the star enters and
leaves the instability region.}
\end{figure}
Fig.~\ref{fig:frequency-evo} shows the evolution of the angular
velocity as a function of time. As predicted by eq.~(\ref{eq:spindown-time-NS}),
the spindown time increases strongly with decreasing saturation amplitude.
The points show that the semi-analytic results for this time again
agree with the numerical solution%
\footnote{Note that he spindown overshoots at large amplitude since it takes
time for the amplitude to decay and the spindown continues during
this period, as can be seen from the lowest curve which drops below
the dot so that the final frequency lies somewhat below the instability
region. This is mainly an artifact of the simplified model which keeps
the amplitude constant until the instability boundary is reached.
In reality with an explicit saturation mechanism the amplitude has
to vanish at the boundary in the static limit, so it must start decreasing
inside the instability region, see \cite{Alford:2011pi}.%
}. Here we are mainly interested in the spindown aspects and only note
that as far as the thermal evolution is concerned the r-mode effectively
causes a delay in the cooling \cite{Alford:2012jc}. This is significant
initially, but, since the cooling slows down strongly, it is irrelevant
at much later times $t\gg t_{sd}$.

Table~\ref{tab:results} lists the semi-analytic results for a range
of saturation amplitudes for two different neutron stars - the previously
discussed standard star with a canonical mass of $1.4\, M_{\odot}$
and, motivated by the recent observation \cite{Demorest:2010bx},
also a heavy $2.0\, M_{\odot}$ star. Strikingly the final frequency
proves nearly independent of the mass, and is far less sensitive than
eq.~(\ref{eq:final-frequency-NS}) naively suggests. This is apparently
caused by cancellations due to the implicit mass dependence of the
radius and of the parameters $\tilde{J}$, $\tilde{S}$ and $\tilde{L}$.
Despite the explicit normalizations in table~\ref{tab:Integral-parameters.}
these parameters still have a significant implicit dependence on $M$
and $R$ via the density profile of the star. Correspondingly, whereas
the dependence on microscopic material properties is explicitly given
by the semi-analytic expressions eqs.~(\ref{eq:final-frequency})
to (\ref{eq:spindown-time-NS}), the dependence on star properties
like the mass and radius is not properly reflected by the explicit
dependencies alone but the additional implicit dependence via the
averaged quantities in table~\ref{tab:Integral-parameters.} is crucial.
The initial cooling time before the star enters the instability region
is amplitude independent and yields for a star spinning at the Kepler
frequency the small values $t_{sc}=96.4\,\left(55.1\right)\,\mathrm{ms}$,
which rise to a few hours for frequencies at the lower end of the
instability region, which is indeed negligible compared to the long
spindown times. The left part of table~\ref{tab:semi-analytic-results}
finally compares the spindown observables for different stars at a
fixed amplitude $\alpha_{{\rm sat}}=1$. We see that, when direct
Urca processes are allowed, the final temperature is significantly
lower due to the enhanced neutrino cooling. Since the instability
region shrinks at lower temperatures the final frequency is higher
and the spindown time is correspondingly shorter. 

\begin{table}
\begin{tabular}{|c|c|c|c|c|}
\hline 
$\alpha_{sat}$ & $T_{f}\,\left[10^{8}\,\mathrm{K}\right]$ & $\Omega_{f}/\Omega_{K}$ & $f_{f}\,\left[\mathrm{Hz}\right]$ & $t_{sd}\,\left[\mathrm{y}\right]$\tabularnewline
\hline 
$1$ & $12.6\,\left(13.9\right)$ & $0.064\,\left(0.049\right)$ & $61.4\,\left(60.3\right)$ & $12.3\,\left(9.4\right)$\tabularnewline
\hline 
$0.1$ & $8.02\,\left(8.88\right)$ & $0.073\,\left(0.056\right)$ & $69.5\,\left(68.4\right)$ & $582\,\left(445\right)$\tabularnewline
\hline 
$10^{-2}$ & $5.11\,\left(5.66\right)$ & $0.082\,\left(0.063\right)$ & $78.8\,\left(77.5\right)$ & $2.75\,\left(2.10\right)\cdot10^{4}\,$\tabularnewline
\hline 
$10^{-3}$ & $3.26\,\left(3.61\right)$ & $0.093\,\left(0.072\right)$ & $89.3\,\left(87.8\right)$ & $1.30\,\left(0.99\right)\cdot10^{6}\,$\tabularnewline
\hline 
$10^{-4}$ & $2.08\,\left(2.30\right)$ & $0.106\,\left(0.081\right)$ & $101.2\,\left(99.5\right)$ & $6.11\,\left(4.68\right)\cdot10^{7}$\tabularnewline
\hline 
\end{tabular}

\caption{\label{tab:results}Dependence of the results obtained from the semi-analytic
expressions for the characteristic quantities eqs.~(\ref{eq:final-temperature-NS}),
(\ref{eq:final-frequency-NS}) and (\ref{eq:spindown-time-NS}) of
the spindown evolution for neutron star with an APR equation of state
on the saturation amplitude $\alpha_{{\rm sat}}$. We give in each
case results for both a standard $1.4\, M_{\odot}$ star and a heavy
$2.0\, M_{\odot}$ star next to it in parentheses. At the smallest
amplitudes the given spindown times $t_{sd}$ should not be relevant
since they would be so long that the neglected magnetic spindown should
dominate and spin down the star on much shorter time scales.}
\end{table}

\subsubsection{Saturation model dependence}

Let us now discuss the influence of different saturation models $\alpha_{{\rm sat}}\!\left(T,\Omega\right)$.
Since in the general case eq.~(\ref{eq:alpha-ansatz}) the amplitude
varies over the instability region and the coefficient $\hat{\alpha}_{{\rm sat}}$
is a dimensionful parameter, we have to find analogous parameter values
for $\hat{\alpha}_{{\rm sat}}$ to compare saturation models with
different values of $\beta$ and $\gamma$ among each other and with
the constant model discussed so far. Since the evolution spends most
of the time in the final stages of the spindown evolution it is natural
to choose $\hat{\alpha}_{{\rm sat}}$ so that different models have
a similar amplitude in this stage. If we compare models defined by
$\hat{\alpha}_{{\rm sat}}$, $\beta$ and $\gamma$ that have the
same amplitude $\alpha_{{\rm sat}}$ at the end of the evolution,
then they leave the instability region at the same point given by
eqs.~(\ref{eq:final-frequency}) and (\ref{eq:final-temperature}).
This can be easily seen, noting that the conditions $\tau_{S}=\tau_{G}$
and $P_{S}=P_{G}$ that determine the endpoint depend only on the
saturation amplitude $\alpha_{sat}\!\left(T_{f},\Omega_{f}\right)$
at this point. However, the path within the instability region, determined
by the steady state curve eq.~(\ref{eq:steady-state-curve}), and
in particular the thermal behavior will depend on $\beta$ and $\gamma$. 

\begin{figure}
\includegraphics{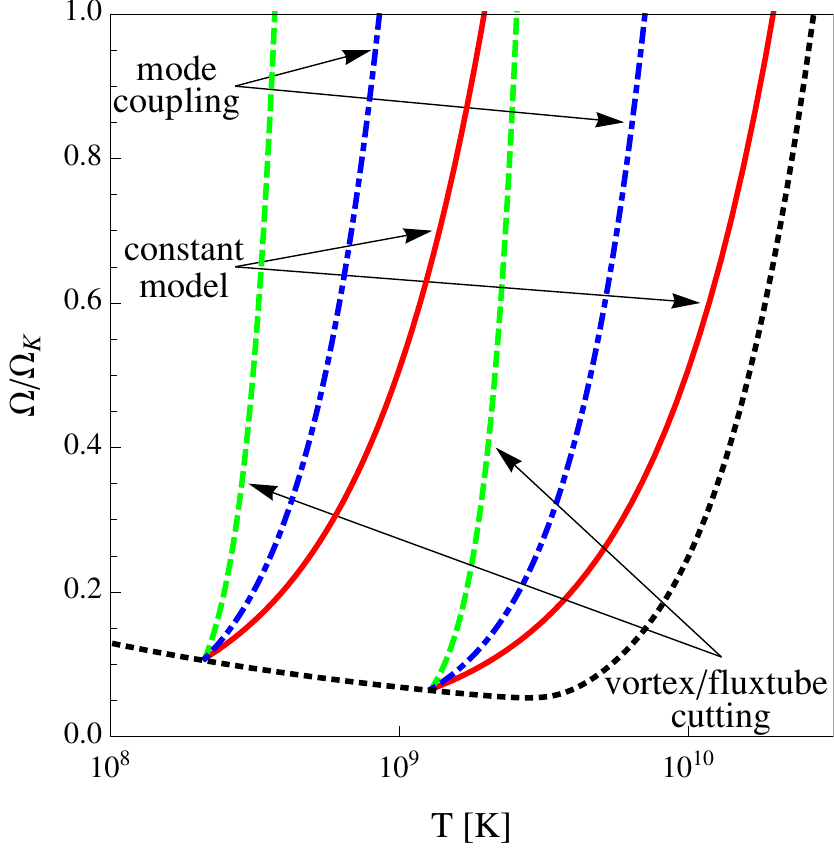}

\caption{\label{fig:saturation-model-dependence}The dependence of the steady-state
curve eq.~(\ref{eq:steady-state-curve}) on the saturation model
for two different constant saturation amplitudes $\alpha_{{\rm sat}}\left(T_{f},\Omega_{f}\right)=10^{-4}$
(left curves) and $1$ (right curves) at the end of the evolution.
The solid line shows the model with constant amplitude $\beta=\gamma=0$
\cite{Owen:1998xg}, the dot-dashed curve shows a mode-coupling model
where $\beta=-4/3$ and $\gamma=-2/3$ \cite{Bondarescu:2013xwa}
and the dashed line denotes the case $\beta=0$, $\gamma=-3$ corresponding
to saturation by cutting of superconducting flux tubes by superfluid
vortices \cite{Haskell:2013hja}.}
\end{figure}

Fig.~\ref{fig:saturation-model-dependence} shows a few examples
of different saturation models for two different saturation amplitudes
$\alpha_{{\rm sat}}=10^{-4}$ and $\alpha_{{\rm sat}}=1$ at the end
of the evolution. The solid lines are the constant model $\beta=\gamma=0$
\cite{Owen:1998xg} discussed above. The dashed curves show the case
of mode coupling with damping of the daughter modes due to shear viscosity
in a star with an impermeable crust \cite{Bondarescu:2013xwa}. In
this case $\beta=-4/3$ and $\gamma=-2/3$ but depending on the particular
damping mechanism of the daughter modes other dependencies are possible
\cite{Bondarescu:2013xwa}. Since the amplitude is lower at large
frequency the dissipative heating is smaller and the steady state
curve becomes steeper and the equilibrium is reached at lower temperatures.

The final, and somewhat extreme, example is the recently proposed
dissipation due to the cutting of superconducting flux tubes by superfluid
vortices when both neutrons and protons pair in a neutron star core
\cite{Haskell:2013hja}. In this case $\beta=0$, $\gamma=-3$. This
moves the steady state curves to even lower temperatures. As discussed
before, this gives a braking index of one so that the spindown is
not power-like but logarithmic and strongly slowed down. In conclusion,
different saturation mechanisms can change both the thermal evolution
and the spindown evolution significantly and generically increase
the spindown time. However, eq.~(\ref{eq:final-frequency}) shows
that the observed insensitivity to the microscopic parameters is not
strongly altered by the saturation model for realistic values $\left|\beta\right|,\left|\gamma\right|\ll\theta$.

\subsection{Comparison to pulsar data}

The semi-analytic expressions finally allow us to compare our theoretical
results to observational data. In most analyses, e.g.\ for low mass
X-ray binaries \cite{Ho:2011tt,Haskell:2012}, the observed pulsar
frequencies and temperatures are compared to numerical computations
of the r-mode instability region. Since the instability boundary is
strongly temperature dependent, information on the temperature is
required to decide if the star is actually inside the instability
region. However, the temperature is unknown for most young pulsars.
For stars which spin slow enough that they are clearly below the minimum
of the instability region the knowledge of the frequency is enough
to rule out that r-mode oscillations are present. However, there is
at least one young star which lies above the minimum of the instability
region. Therefore, we will here go beyond such a static analysis
and take into account the dynamic evolution as well as the frequency
eq.~(\ref{eq:final-frequency}) where the star actually leaves the
instability region.

\begin{figure}
\includegraphics[scale=0.97]{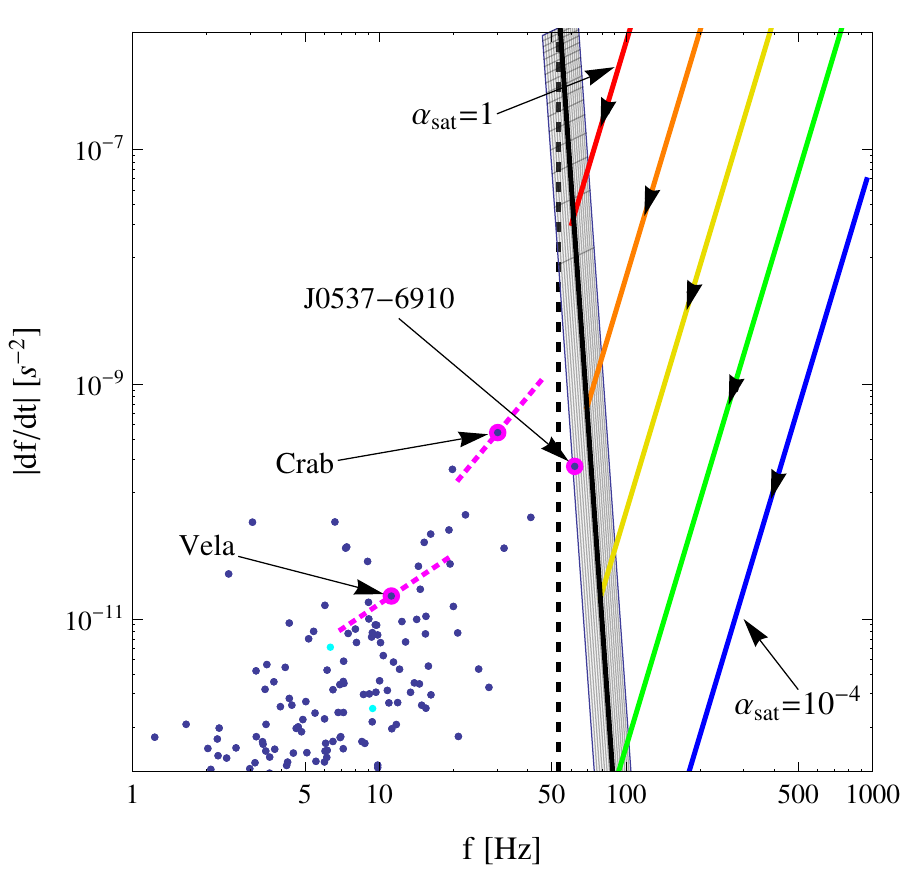}

\caption{\label{fig:F0F1data}The spindown evolution of a young neutron star
compared to observed pulsar data from the ATNF catalog \cite{Manchester:2004bp}.
The solid lines show the evolution of the $1.4\, M_{\odot}$ star
for different saturation amplitudes ranging from $\alpha_{{\rm sat}}=1$
(top) to $\alpha_{{\rm sat}}=10^{-4}$ (bottom). The vertical dashed
line to the left shows the lower frequency boundary of the instability
region. The steep solid line is formed by the endpoints of the evolution
for different values of $\alpha$ and thereby represents the boundary
of the instability region in $f$-$\dot{f}$-space, whereas the gray
band reflects the error due to the uncertainty in the underlying parameters.
The dotted line segments show the current evolution for two stars
for which a reliable braking index is available \cite{1983bhwd.book.....S,Lyne:1996}. }
\end{figure}

In addition to the spin frequencies also the spindown rates are known
for many young stars. Naturally, this additional information provides
a stronger constraint on spindown models. The time derivative of the
frequency is readily obtained from eq.~(\ref{eq:Omega-equation-saturation}).
Fig.~\ref{fig:F0F1data} compares the spindown evolution for different
saturation amplitudes to the data for young stars that still have
a significant spindown rate. As can be seen there is one pulsar J0537-6910
that spins fast enough that it could be within the instability region
of a standard neutron star whose minimum frequency is denoted by the
vertical dotted line to the left%
\footnote{The analytic result for the minimum of the instability region given
in \cite{Alford:2010fd} would allow us to estimate also the uncertainties
on the minimum frequency, but we refrain from this for better readability.%
}. Taking into account the dynamical evolution, the additional knowledge
of the spindown rate alleviates the uncertainty with respect to the
saturation amplitude, since larger amplitudes lead to significantly
larger spindown rates. This is shown by the black line which is formed
by the semi-analytic results for the endpoints of the evolution and
which represents the boundary of the instability region in $f$-$\dot{f}$-space.
The gray band reflects the uncertainty in the underlying parameters
eq.~(\ref{eq:frequency-uncertainty}). The fastest pulsar J0537-6910
is obviously compatible with being outside of the instability region.
We conclude that no observed star lies definitely inside the instability
region (see also \cite{Arras:2002dw}) which confirms the r-mode scenario
for a sufficiently large saturation amplitude resulting in a short
spindown time. Moreover, J0537-6910 lies clearly below the line for
our APR neutron star and just at the lower edge of the uncertainty
band. Therefore, very likely no young star observed so far currently
emits gravitational waves because of the r-mode instability. This
conclusion is strengthened by the absence of gravitational wave signals
from observed young pulsars - including J0537-6910 - in the latest
LIGO data \cite{Collaboration:2009rfa}. Unfortunately, the measured
braking index of $n\!\approx\!-1.5$ for J0537-6910 \cite{2006ApJ...652.1531M}
seems to be strongly affected by the glitch activity of the pulsar
\cite{1983bhwd.book.....S} and therefore cannot provide conclusive
evidence on the current spindown mechanism. For other pulsars the
measured braking indices%
\footnote{The braking index (see eq.~(\ref{eq:generic-spindown-model})) would
be $3$ for pure magnetic dipole emission, $5$ for spindown via gravitational
waves due to an ellipticity of the star \cite{1983bhwd.book.....S}
and a value $\leq7$ for spindown due to r-modes as discussed above.%
} $n\lesssim3$, as shown in fig.~\ref{fig:F0F1data} for the Crab
\cite{1983bhwd.book.....S} and the Vela \cite{Lyne:1996} pulsar,
already revealed a dominant electromagnetic spindown. For the Crab
pulsar this was further strengthened by the LIGO S5 limit \cite{Abbott:2008fx},
showing that it could at most emit a few percent of the rotational
energy loss as gravitational waves. 

The remarkable aspect of eq.~(\ref{eq:final-frequency-NS}) for the
final frequency of the spindown is that it is a robust lower bound
and is not affected by our ignorance of the star's composition. Although
our analysis was performed for a standard neutron star, any additional
exotic forms of matter entail enhanced damping so that their instability
regions do not extend to such low frequencies. The insensitivity of
the final frequency eq.~(\ref{eq:final-frequency}) to the underlying
parameters, therefore provides a quantitative, universal lower bound
given by eq.~(\ref{eq:frequency-uncertainty}). As a consequence
our result shows that the r-mode picture of the spindown of young
pulsars presents a viable explanation for the low spin rates of young
pulsars if the saturation amplitude is large enough $\alpha_{{\rm sat}}\gtrsim0.01$
to spin down the pulsar in a time $t_{sd}<10^{4}\,{\rm y}$. An upper
limit for the saturation amplitude $\alpha_{{\rm sat}}<1$ comes from
the observed timing data of J0537-6910, as noted in \cite{Owen:2010ng},
since r-modes would have otherwise spun it down to its current frequency
in less than a hundred years which is much shorter than its age estimate
\cite{Wang:1998}. Whereas the r-mode picture yields a direct and
quantitative explanation for the maximum observed spin frequency,
other mechanisms, like the possibility that all stars are already
born with lower frequencies or the subsequent spindown via magnetic
dipole radiation, generally only provide probabilistic answers why
fast stars with a small initial magnetic dipole moment resulting in
a slow spindown could not form. Despite the general trend that according
to the dynamo mechanism fast spinning stars feature larger magnetic
fields \cite{Bonanno:2006su}, the required spindown evolution studies
rely on Monte Carlo population synthesis modeling and involve uncertainties
that are to some extent exploited to reproduce the statistical aspects
of the observed pulsar distribution \cite{FaucherGiguere:2005ny}.
Finally, further indirect evidence for the above r-mode bound comes
from pulsars in old and very stable double neutron star binaries which
are not recycled and correspondingly should not have experienced any
appreciable spin-up during their evolution. The fastest of these,
J0737-3039A located in the only known double pulsar system \cite{Lyne:2004cj},
has a frequency of $f\approx44\,{\rm Hz}$ which is still intriguingly
close to the quantitative r-mode bound eq.~(\ref{eq:frequency-uncertainty}).

\section{Gravitational wave emission}

As discussed in the previous section, all presently observed young
pulsars are likely outside of the instability region and so should
not emit gravitational radiation due to r-modes. Nevertheless, it
is possible that in the future we may observe a young pulsar that
is still in the unstable region. Furthermore, gravitational waves
might even be detected from so far unknown young compact stars that
have not been seen because their electromagnetic radiation is either
too faint or it is absorbed or outshined by the supernova remnant.
This possibility is interesting since there are several known young
supernova remnants, including most notably SN 1987A, where a compact
object has not been observed, yet. In view of the upcoming next generation
of gravitational wave detectors like advanced LIGO \cite{Harry:2010zz}
it is therefore important to understand the expected gravitational
radiation of such sources in detail. 

Above we already gave semi-analytic expressions for two important
quantities in this respect, namely the rotational angular velocity
$\Omega_{f}$ (eq.~(\ref{eq:final-frequency})) when the star exits
the r-mode instability region, which is related to the lower bound
$\nu_{f}$ of the gravitational wave spectrum via $\nu=2/\left(3\pi\right)\Omega$,
and the spindown time $t_{sd}$ (eq.~(\ref{eq:spindown-time})),
which is the maximum age up to which gravitational waves from such
a source could be observed. For a standard neutron star, using eq.~(\ref{eq:frequency-uncertainty}),
we estimate $\nu_{f}\approx\left(81.9\pm12.5\right)\mathrm{Hz}\,\,\alpha_{{\rm sat}}^{-5/92}$.
To have a reliable estimate for the lower bound of the spectrum is
particularly important since advanced LIGO will have a significantly
improved sensitivity below 100 Hz which coincides with the final stages
of the r-mode spindown. Since the r-mode evolution becomes slow at
low frequencies this region is where the evolution spends the most
time and the analytic expression for the spindown time as a function
of the dependent parameters should then allow us to estimate the probability
that sources within the reach of advanced LIGO are currently active
and observable.

\subsection{Gravitational emission due to r-modes}

In this section we follow and extend the previous analysis \cite{Owen:1998xg}
and derive general results for the gravitational wave strain based
on our analytic solution of the r-mode evolution. An explicit expression
for the gravitational wave strain, averaged over polarizations and
the orientation and position of a source at distance $D$ on the sky,
was given in \cite{Owen:1998xg,Thorne:1980ru}

\begin{equation}
h=\sqrt{\frac{3}{80\pi}}\frac{\omega^{2}S_{22}}{D}\:,\label{eq:strain}
\end{equation}
with the mass current quadrupole moment

\begin{equation}
S_{22}=\sqrt{2}\frac{32\pi}{15}GM\alpha\Omega R^{3}\tilde{J}\:.\label{eq:quadrupole-moment}
\end{equation}
The strain depends both on the time-dependent r-mode amplitude and
angular velocity

\begin{equation}
h\!\left(t\right)=\sqrt{\frac{2^{15}\pi}{3^{5}5^{3}}}\frac{\tilde{J}GMR^{3}\alpha\!\left(t\right)\Omega\!\left(t\right)^{3}}{D}\label{eq:gw-strain}
\end{equation}
This expression has 3 different limits. Initially the amplitude rises
exponentially whereas the frequency is fixed. Depending on the amplitude
there can be an intermediate period where both $\alpha$ is saturated
but $\Omega$ does not appreciably spin down yet, since the first
term in eq.~(\ref{eq:spindown-solution}) dominates. Finally, at
late times, when the second term in eq.~(\ref{eq:spindown-solution})
dominates, the expression becomes independent of the initial conditions.

At saturation the r-mode amplitude can be replaced using the spindown
equation (\ref{eq:Omega-equation-saturation}). which gives

\begin{equation}
h\!\left(t\right)=\sqrt{\frac{9}{20}\frac{GI\dot{\Omega}\!\left(t\right)}{D^{2}\Omega\!\left(t\right)}}\label{eq:spindown-limit-strain}
\end{equation}
This expression gives the strain in terms of the present frequency
and the spindown rate. Assuming that r-modes dominate the spindown,
this expression yields for sources with known timing data the standard
\emph{spindown limi}t. This limit, which is analogously obtained e.g.
for gravitational waves due to an ellipticity of the source, is very
useful since it gives the maximum signal that can be expected for
a given known source. There is only a chance to detect gravitational
waves - or alternatively learn something about the source from a non-detection
- if the detector sensitivity is lower than the spindown limit. 

However, according to the discussion in section \ref{sec:Spindown-of-young}
the observed pulsars do not emit gravitational waves due to r-modes
any more, and timing data is not available for other promising young
sources, like compact objects in supernova remnants or pulsar spin
nebulae. Only the age is usually roughly known. A prediction for the
gravitational strain requires therefore an actual solution of the
evolution. Previously a standard spindown equation with a general
breaking index \cite{1983bhwd.book.....S} has been used to derive
the strain as a function of the age of the source \cite{Wette:2008hg}.
This expression also applies to the emission due to ellipticity and
at late times it proved to be independent of the saturation amplitude.
In the case of r-modes the strong heating generally requires a solution
of the coupled set of evolution equations (\ref{eq:Omega-equation-saturation})
and (\ref{eq:T-equation-saturation}), though. As discussed in section
\ref{sec:Semi-analytic-expressions} the spindown evolution decouples
only in the special case of a constant saturation amplitude. However,
due to the r-mode heating the evolution predicted by this toy model
can be quite different from that obtained from the coupled system
for realistic saturation mechanisms, where the saturation amplitude
is generally a function of temperature and frequency. Our analysis
leading to the effective spindown evolution eq.~(\ref{eq:effective-spindown-equation})
confirms that the study performed in \cite{Wette:2008hg} does indeed
apply to r-mode gravitational wave emission. The effective braking
index eq.~(\ref{eq:effective-braking-index}) gives the explicit
connection for a realistic r-mode saturation mechanism. Inserting
the frequency evolution (eq.~(\ref{eq:spindown-solution})) and the
general form for the saturation amplitude (eq.~(\ref{eq:alpha-ansatz})),
there are significant cancellations and in the late time limit, where
the frequency is far below the initial frequency, the strain becomes

\begin{equation}
h\!\left(t\right)\xrightarrow[\Omega\ll\Omega_{i}]{}\sqrt{\frac{3}{40}\frac{{\cal C}GI}{D^{2}t}}\:.\label{eq:independent-strain}
\end{equation}
with the same constant ${\cal C}$ describing the saturation mechanism
(eq.~(\ref{eq:saturation-model-factor})) that appeared before in
the spindown time. This expression agrees with the result in given
in \cite{Owen:2010ng} using the effective braking index eq.~(\ref{eq:effective-braking-index}).
A striking property of eq.~(\ref{eq:independent-strain}) is that
it is indeed \emph{independent} of the unknown saturation amplitude
and even of the complete microphysics, including the cooling mechanism.
Therefore, the strain of a given source is, up to moderate uncertainties
in the moment of inertia (eq.~(\ref{eq:moment-of-inertia-bounds}))
and the constant factor ${\cal C}$ involving $\beta$ and $\gamma$,
completely determined by its distance and its age. We recall that
the constant ${\cal C}$ is $1$ for a constant saturation model ($\beta\!=\!\gamma\!=\!0$),
where the expression directly reduces to the result given in \cite{Wette:2008hg,Owen:2010ng},
and somewhat larger for general saturation models. Because of the
square root, the dependence on the saturation model is even weaker
and the result does not change by more than some tens of per cent
for realistic values $\left|\beta\right|,\left|\gamma\right|<2$ predicted
by different saturation mechanisms%
\footnote{Although this is not the case for the saturation mechanisms cited,
a general saturation mechanism, e.g. non-linear hydrodynamics effects
\cite{Lindblom:2000az,Kastaun:2011yd}, could impose a more complicated
function $\alpha_{{\rm sat}}\!\left(\Omega,T\right)$. As argued before
such a function is nevertheless approximated locally in different
regions by the simple power law form eq. (\ref{eq:alpha-ansatz}).
The prefactor in eq.~(\ref{eq:independent-strain}) would then be
locally different, but because of the small size of the variation
it would not change the conclusion.%
} \cite{Arras:2002dw,Bondarescu:2008qx,Bondarescu:2013xwa,Alford:2011pi}.
Therefore, the simple expression eq.~(\ref{eq:independent-strain})
confirms and generalizes the result of \cite{Wette:2008hg,Owen:2010ng}
to the more complicated case of r-mode emission with realistic saturation
mechanisms, where the thermal heating significantly affects the spindown.

\begin{figure}
\includegraphics{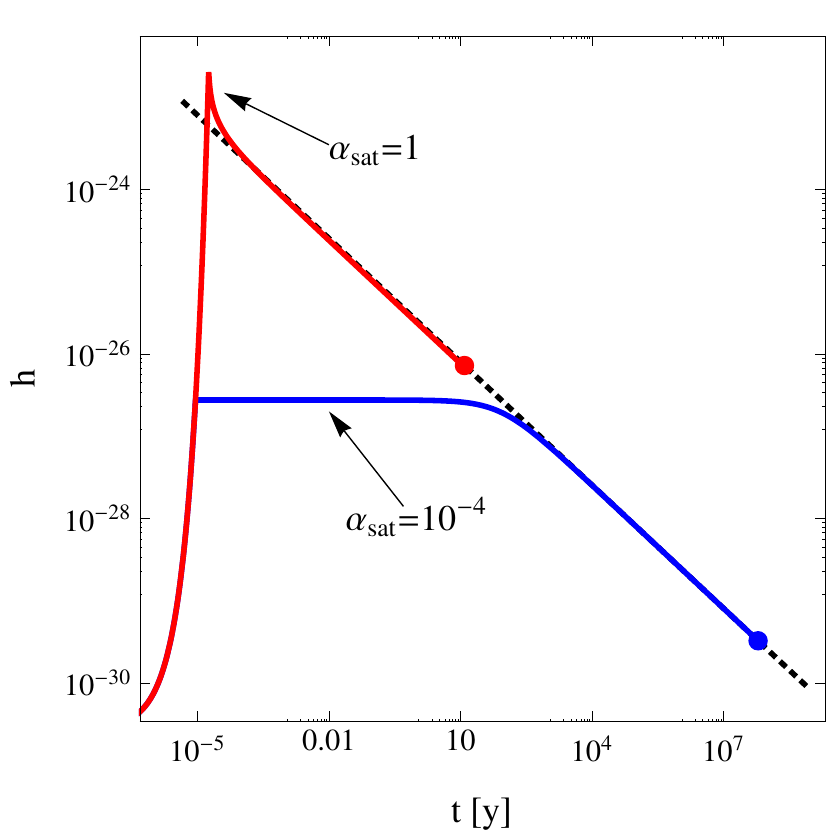}

\caption{\label{fig:gravitational-wave-strain}The time evolution of the gravitational
wave strain of a $1.4\, M_{\odot}$ APR neutron star, spinning initially
with its Kepler frequency and located at a distance of 1 Mpc. We show
the numerical curves for two different r-mode saturation amplitudes
$\alpha_{{\rm sat}}=1$ (top) and $\alpha_{{\rm sat}}=10^{-4}$ (bottom).
The dotted line denotes the asymptotic late time expression eq. (\ref{eq:independent-strain}).
The dots denote the analytic result for the minimum strain amplitude
of the gravitational radiation emitted at the end of the r-mode evolution
and the endpoints for other amplitudes lie on the dotted line.}
\end{figure}

The analytic expression for the cutoff frequency eq.~(\ref{eq:final-frequency})
also allows us to determine the final gravitational wave strain emitted
in the last stage of the evolution $h_{f}\equiv h\!\left(\Omega_{f}\right)$
in terms of the final frequency $\Omega_{f}$ eq.~(\ref{eq:final-frequency}).
For a neutron star the final gravitational strain is

\begin{equation}
h_{f}^{\left(NS\right)}\approx7.23\cdot10^{-27}\alpha_{{\rm sat}}^{\frac{77}{92}}\left(\frac{\mathrm{Mpc}}{D}\right)\:.\label{eq:minimum-strain-NS}
\end{equation}
The emitted gravitational wave strain during the evolution is shown
in fig.~\ref{fig:gravitational-wave-strain} for different values
of the saturation amplitude. The numerical curves are compared to
the analytic results for the minimum gravitational strain, which is
emitted at the end of the evolution eq.~(\ref{eq:minimum-strain-NS})
and denoted by the points. As can be seen, small amplitude r-modes
have a lower maximum amplitude but emit low amplitude gravitational
waves over longer times. The plot shows that given the relevant age
range of young pulsars, for saturation amplitudes $\alpha_{sat}\gtrsim10^{-4}$
the late time approximation eq.~(\ref{eq:independent-strain}) should
be justified. For much smaller amplitudes the star is hardly spun
down by r-modes, but in this case other spindown mechanisms, like
magnetic dipole radiation, will dominate and determine the spindown
$\Omega\!\left(t\right)$. This scenario, which is unlikely according
to our results in section \ref{sec:Spindown-of-young}, is discussed
for completeness in appendix \ref{sec:Very-low-saturation}. 

\subsection{Signal detectability}

The comparison of signals to the detector sensitivity is generally
done in terms of the \emph{intrinsic strain amplitude}%
\footnote{This is the amplitude measured by a detector at one of the earths
poles of a source directly over it whose rotation axis is parallel
to that of the earth.%
} $h_{0}$ which is in case of r-modes given by \cite{Owen:2010ng}

\begin{equation}
h_{0}=\sqrt{\frac{8\pi}{5}}\frac{\tilde{J}GMR^{3}\left(2\pi\nu\right)^{3}\alpha_{{\rm sat}}}{D}=\sqrt{\frac{25}{3}}h\:.\label{eq:intrinsic-strain-amplitude}
\end{equation}
Using the general constraints on the moment of inertia eq.~(\ref{eq:I-tilde-bounds})
and (\ref{eq:moment-of-inertia-bounds}) and the range of different
saturation models \cite{Arras:2002dw,Bondarescu:2008qx,Bondarescu:2013xwa,Alford:2011pi,Owen:1998xg}
yields for a constant saturation amplitude the remarkably definite
result
\begin{align}
h_{0} & \approx2.3_{-0.8}^{+3.5}\times10^{-27}\sqrt{\frac{1000\,{\rm y}}{t}}\frac{1\,{\rm Mpc}}{D}\:,\label{eq:numeric-intrinsic-strain}
\end{align}
where the uncertainty limits are conservative and should, as discussed
before, likely exceed the realistic range.

For an actual detection of gravitational waves a systematic search
strategy is required which takes into account the periodicity of the
signal. In particular, this requires a detailed model of the expected
signal and its change over the observation interval in order to search
for it in the data via a given statistical method, like e.g.\ matched
filtering. Only for known isolated pulsars with precise timing data,
like frequency and spindown rate, is such detailed knowledge of the
phase-coherent signal available. Otherwise a huge parameter space
for whole classes of models with different frequencies, spindown rates,
... has to be searched which greatly complicates the analysis%
\footnote{This holds even more for sources in binaries, which we will not discuss
here since the fraction of young pulsars in such systems is small.%
}. Reducing the computational cost in such searches is crucial since
a coherent search over time scales of order years for such an unconstrained
system is still not manageable. Therefore, one important aim of our
work is to narrow down the parameter space and to identify the most
promising parameter regions for different sources. 

Naturally, the likelihood of detecting a continuous signal in gravitational
wave searches increases with the observation time. This is usually
taken into account by giving the equivalent intrinsic strain amplitude
that would be observed within a given search at 95\% confidence level.
For a coherent search the intrinsic strain amplitude is given by \cite{Abbott:2003yq}

\begin{equation}
h_{0}^{95\%}=\Theta\sqrt{\frac{S_{h}\!\left(f\right)}{\Delta t}}\:.\label{eq:observational-intrinsic-strain}
\end{equation}
It is a property of the particular search analysis and depends both
on the power spectral density%
\footnote{The ``power spectral density'' is just the square of the amplitude
spectral density of the noise expressed in equivalent gravitational
wave strain, as given e.g.\ by the LIGO collaboration \cite{Abbott:2007kv}.%
} $S_{h}$ of the detector noise and the observation time $\Delta t$.
The equivalent intrinsic strain amplitude eq.~(\ref{eq:observational-intrinsic-strain})
relies on matched filtering and requires knowledge of the detailed
form and time evolution of the coherent signal. Since this might not
always be available, the sensitivity obtained using this method is
an optimistic estimate for other methods, where e.g.~signals are
combined incoherently. For observed isolated pulsars with known timing
behavior the prefactor is $\Theta\approx11.4$ and for known isolated
sources without timing data, like Cas A or hypothetical localized
compact sources within known supernova remnants, it is about 35 \cite{Wette:2008hg}.
In the following we will discuss the detectability of gravitational
waves in terms of the intrinsic strain amplitude, but other sensitivity
measures have been use in the literature \cite{Owen:1998xg} and we
will discuss some of them and their reliability in appendix \ref{sec:Alternative-sensitivity-measures}.

\begin{figure*}
\includegraphics[scale=0.85]{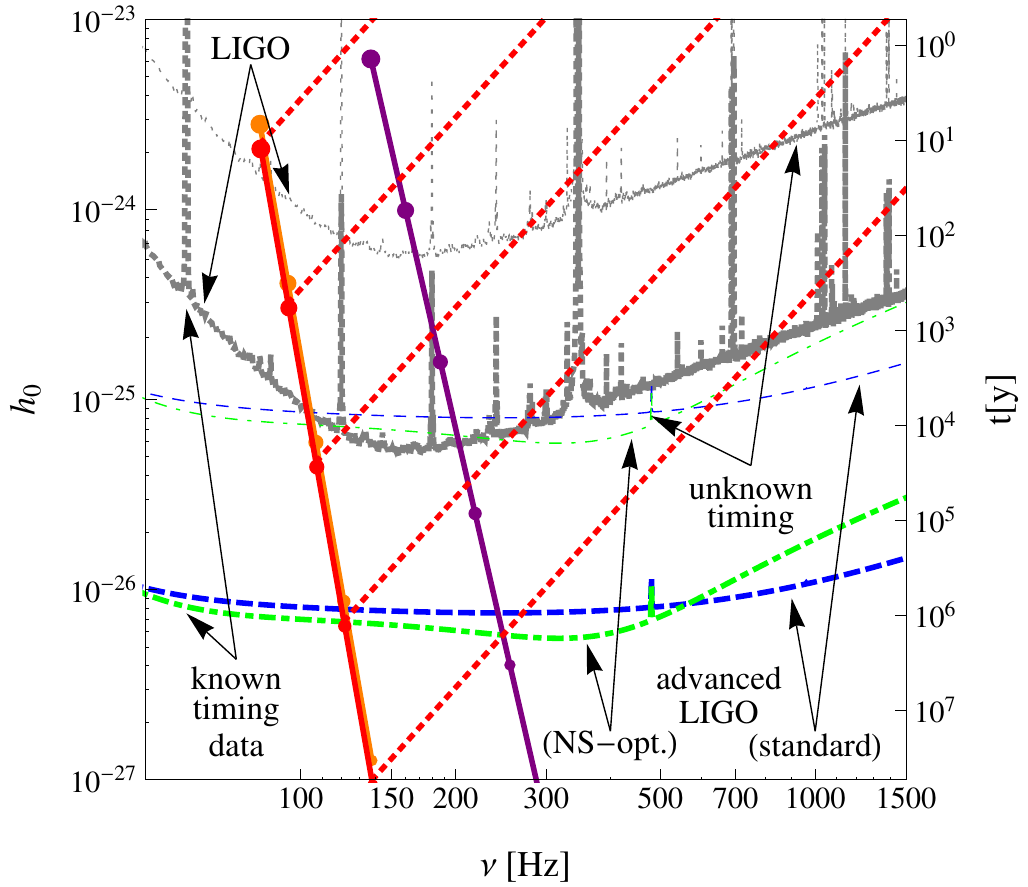}\includegraphics[scale=0.85]{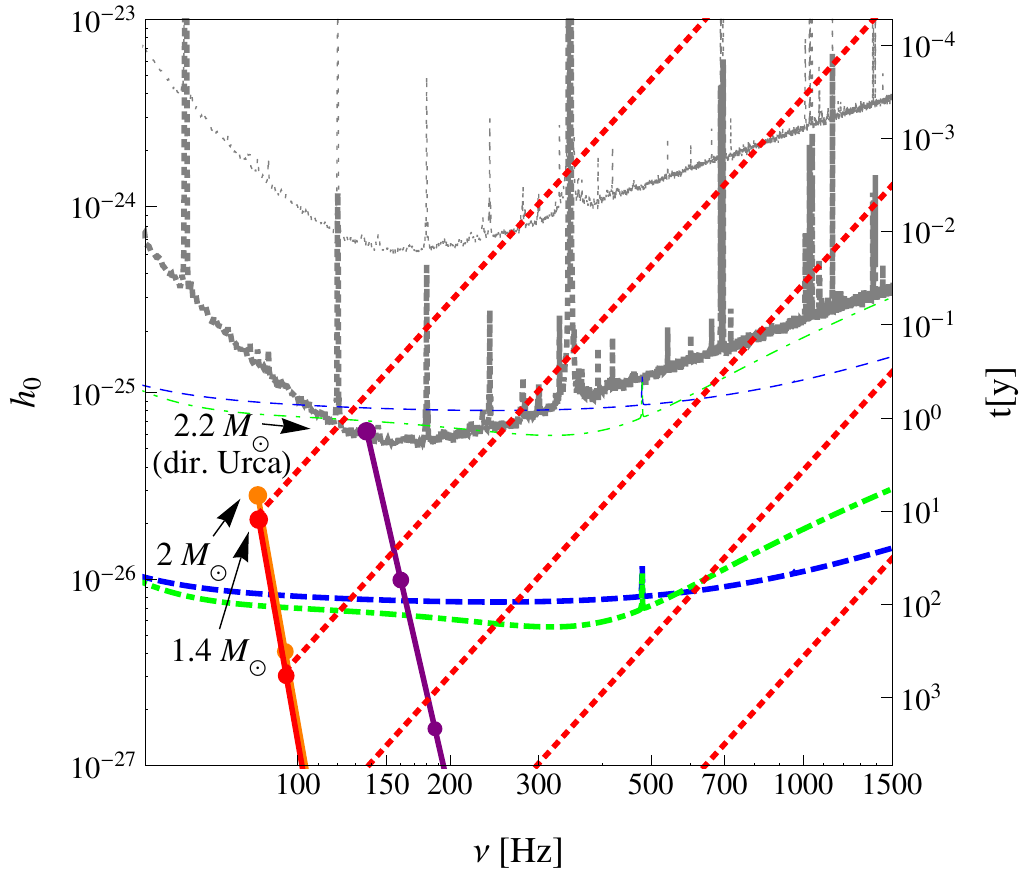}

\caption{\label{fig:intrinsic-strain}The frequency dependence of the intrinsic
strain amplitude for different sources compared to the signal detectability
of different detectors and for different observation periods. \emph{Left}:
Source within the galaxy (10 kpc), \emph{right}: source within the
local group (1 Mpc). The steeply decreasing solid lines show the minimum
strain at the end of the r-mode evolution as a function of the gravitational
wave frequency. These are shown for the three different APR stars
given in tab.~\ref{tab:parameter-values}. The dots show the strain
for a few fixed values of the saturation amplitude of different orders
of magnitude $\alpha_{{\rm sat}}=1$ (top) to $\alpha_{{\rm sat}}=10^{-4}$
(bottom), and the dotted curves ending at these points show for the
$1.4\, M_{\odot}$ star the evolution at earlier times. These curves
are given for the constant r-mode saturation model which is a lower
bound for the strain amplitude within other saturation mechanisms
(see eq.~(\ref{eq:independent-strain})). The theoretical results
are compared to the sensitivity of LIGO (solid) and the anticipated
sensitivity of advanced LIGO \cite{Harry:2010zz} in the standard
mode (dashed) and the neutron star optimized configuration (dot dashed).
These are given both for a known pulsar search with one year of data
(thick) and a search for potential sources without timing information
using one month of data (thin). The right labels denote the age of
the source assuming a constant saturation amplitude.}
\end{figure*}

In fig.~\ref{fig:intrinsic-strain} our results for $h_{0}$ from
gravitational wave emission by r-modes for different neutron star
models are compared to the sensitivity curves%
\footnote{The sensitivity curves have been released by the LIGO scientific collaboration
and were obtained from http://www.ligo.org/science/data-releases.php
and as technical document T0900288 at http://dcc.ligo.org.%
} for the LIGO detector (S6 run) and the projected advanced LIGO sensitivity
\cite{Harry:2010zz}, in both the standard (``ZERO\_DET\_high\_P'')
and the neutron star optimized mode (``NSNS\_Opt''). For each detector,
we show the two exemplary cases of a coherent search for known pulsars
(where $\Theta=11.4$) using one year of data (thick) and a search
for sources without timing data, i.e.\ neutron stars or unseen compact
remnants (using $\Theta=35$) with an analysis of a month of coherent
data (thin). The left panel assumes a typical source within the galaxy
($D=10\,{\rm kpc}$) and the right panel a source in the local group
of galaxies ($D=1\,{\rm Mpc}$). The label on the right axis also
shows the age of the object according to the direct relation for constant
saturation amplitude eqs.~(\ref{eq:independent-strain}) and (\ref{eq:intrinsic-strain-amplitude}).
Let us first discuss this general relation. As can be seen LIGO could
have detected r-mode emission from known galactic pulsars up to an
age of about ten thousand years. According to our analysis in the
preceding section, this should have been sufficient to see gravitational
wave emission from all such sources that can be seen, but unfortunately
there are likely no known young pulsars that are currently in the
r-mode phase. For known pulsars the sensitivity enhancement of advanced
LIGO might therefore not help. But in case we detect in the future
an even faster pulsar that is less than a hundred years old, then
its r-mode emission could be detected with advanced LIGO even if it
is extragalactic. There are a few such candidates including most notably
SN 1987A in the Large Magellanic Cloud. For sources without timing
data LIGO would have only seen galactic sources that are at most a
few dozens of years old and only within a narrow frequency range.
There has not been a galactic supernova for a long time and therefore
it is not surprising that LIGO did not make a detection. In contrast
advanced LIGO could, thanks to its flat noise spectrum, detect a galactic
source over the entire relevant frequency range and up to an age of
more than a thousand years. As discussed before, this is exactly the
age range favored by the r-mode scenario for the spindown of the observed
pulsars and therefore advanced LIGO is a very promising machine for
the detection of gravitational waves from r-mode oscillations of young
neutron stars. Moreover, for extragalactic supernovae the sensitivity
of advanced LIGO should be sufficient to see r-mode oscillations of
a compact remnant in the immediate aftermath up to about a year. For
sources as far as the Virgo cluster of galaxies, as originally discussed
in \cite{Owen:1998xg}, even with advanced LIGO a detection should
only be possible under extremely lucky circumstances and third-generation
detectors like the Einstein telescope \cite{Punturo:2010zz} will
be necessary to take advantage of the much larger number of more distant
sources. 

Next consider our theoretical results in fig.~\ref{fig:intrinsic-strain},
which are given for the different neutron stars listed in table \ref{tab:parameter-values}.
Namely the $1.4\, M_{\odot}$ and $2\, M_{\odot}$ APR stars for which
only modified Urca reactions are possible and the $2.2\, M_{\odot}$
star where direct Urca reactions are present within its core. The
steeply descending solid lines show the minimal signal emitted at
the end of the r-mode spindown evolution (where the star spends most
of its time) as a function of the frequency, which in turn is a function
of the unknown saturation amplitude. The dots show the result for
fixed saturation amplitudes of different orders of magnitude ranging
from $\alpha_{{\rm sat}}=1$ (top) to $\alpha_{{\rm sat}}=10^{-4}$
(bottom). The dotted curves show the r-mode evolution for the $1.4\, M_{\odot}$
model for these saturation amplitudes. As expected the signal is larger
and emitted at higher frequencies if the star is younger but a star
also spends far less time in these earlier stages. Although a star
of a given age emits gravitational waves with a fixed strain amplitude,
depending on the unknown saturation amplitude the frequency of these
gravitational waves can be very different. As suggested by the analytic
results, the strain depends little on the mass and the details of
the composition, and the size of (suppressed) error bands on these
curves would be given by the errors in prefactor of eq.~(\ref{eq:numeric-intrinsic-strain}).
This insensitivity holds even for the case of the heavy star where
direct Urca emission is possible. As noted before, because of the
much stronger cooling in this case the star spins down at a lower
temperature where the boundary of the instability region is at a larger
frequency, so that the lower gravitational wave cutoff frequency is
considerably higher. Yet, up to this point the gravitational wave
signal is nearly identical to the standard case with modified Urca
cooling. This is an important observation since the data on Cas A
\cite{Heinke:2010cr}, which is the only young star for which the
cooling has been explicitly observed, clearly points to an enhanced
neutrino emission mechanism. The standard explanation \cite{Page:2010aw,Yakovlev:2010ed}
is emission due to pair breaking of superfluid neutrons which is in
between the cases of modified and direct Urca emission discussed here.

\subsection{Searches for promising sources}

The fact that nearly all known pulsars spin too slowly to emit gravitational
waves due to r-modes at the moment requires us to consider other potential
sources. Here we give the expected signal for a few examples of known
possible sources that could be young enough to emit gravitational
waves due to r-modes. These consist of, in addition to the fastest
pulsar J0537-6910 discussed above, the young neutron star Cassiopeia
A from which no pulsation has been observed \cite{2010ApJ...709..436H}
and several supernova remnants where no compact object has been detected
so far. However, it has recently been suggested that the hard X-ray
emission from SN 1957D results from a pulsar wind nebula and the compact
remnant which produced it would then be the youngest indirectly observed
pulsar \cite{Long:2012}. We take into account both the age and the
available distance estimate for these sources shown in tab.~\ref{tab:source-examples}.
As noted before, the strain amplitude is fixed by these parameters,
but the frequency can vary depending on the unknown r-mode saturation
amplitude.

\begin{figure}
\includegraphics[scale=0.85]{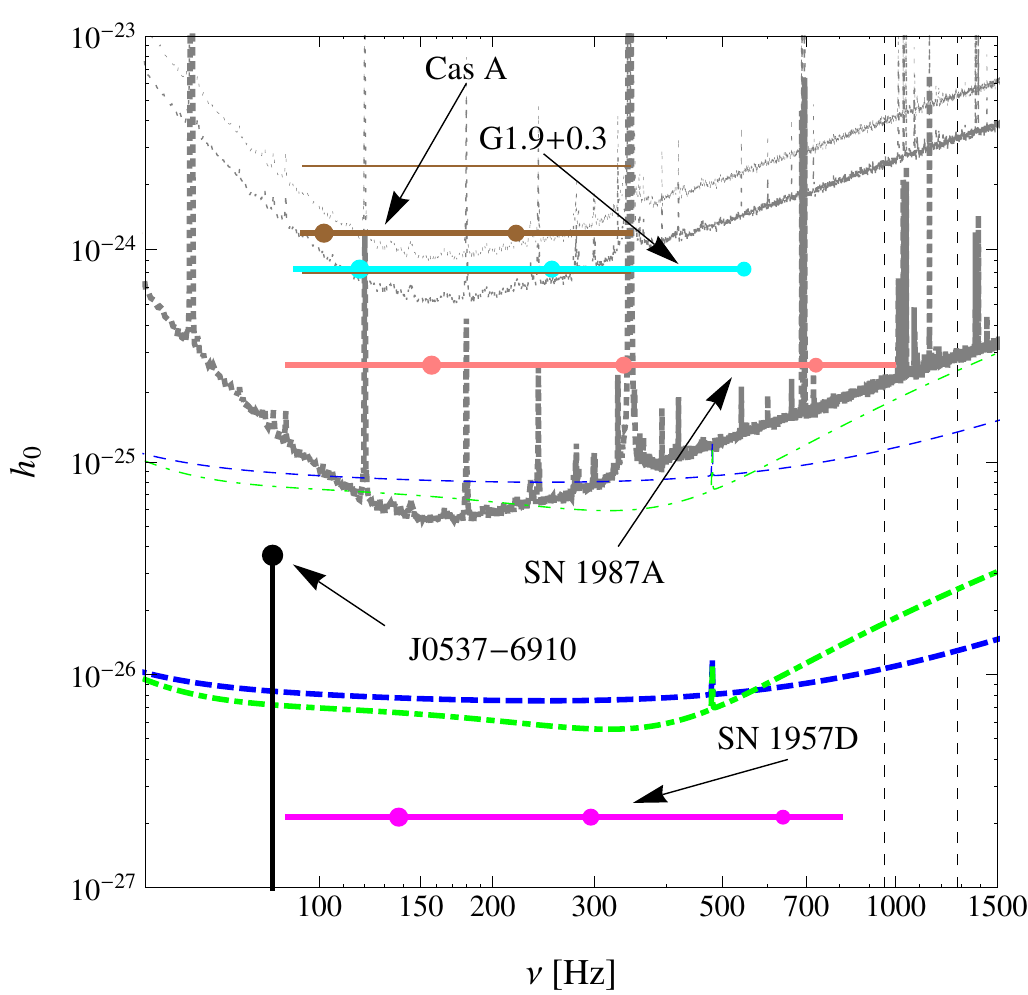}

\caption{\label{fig:strain-sources}The frequency dependence of the intrinsic
strain amplitude for potential realistic sources. The actual age of
these sources as well as their distance have been used and the expected
frequency range determined from these known parameters is shown. The
frequency range corresponds to a range of saturation amplitudes and
the dots denote the corresponding values for different orders of magnitude
from $\alpha_{{\rm sat}}=0.1$ to $10^{-3}$. We also included the
fastest young pulsar J0537-6910 for which the observed frequency is
used and where the dot represents the result if r-mode spindown dominates
and lower values are possible if other spindown sources are important.
The detector curves are the same as in fig.~\ref{fig:intrinsic-strain}
with the addition of the uppermost curve which shows the sensitivity
of a previous Cas A search \cite{Abadie:2010hv}. The two thin vertical
lines show for comparison the corresponding frequency of the fastest
observed (old) pulsar and that corresponding to the maximum Kepler
frequency of the star. }
\end{figure}
Because of the significant computational cost of a gravitational search
it is essential to know the spindown model and the expected timing
parameter range for a given source \cite{Abbott:2007kv,Wette:2008hg}.
In gravitational wave searches the expected signal is usually modeled
by a generic spindown model \cite{1983bhwd.book.....S}

\begin{equation}
\dot{f}\sim-f^{n}\quad\Rightarrow\quad\ddot{f}=n\dot{f}^{2}/f\label{eq:generic-spindown-model}
\end{equation}
with a fixed braking index $n$. Since the spindown evolution is
coupled to the thermal evolution in the case of r-mode emission, up
to now it might have seemed questionable whether the simple model
eq.~(\ref{eq:generic-spindown-model}) is appropriate. However, our
general analysis in section \ref{sec:Semi-analytic-expressions} showed
that the thermal evolution is always faster than the spindown (eq.~(\ref{eq:thermal-spindown-boundary}))
and can be solved in a first step. Therefore the effective spindown
equation (\ref{eq:effective-spindown-equation}) takes indeed the
standard form eq.~(\ref{eq:generic-spindown-model}), yet with an
effective r-mode braking index $n_{rm}$ given in eq.~(\ref{eq:effective-braking-index}).
As discussed, the saturation model can change the braking index from
the canonical value $7$ to significantly smaller values. E.g. for
mode coupling with damping due to shear viscosity \cite{Bondarescu:2013xwa}
the braking index is considerably reduced to $n_{rm}^{\left(mc\right)}=4$.
This is important both for the search for gravitational waves, as
well as for the interpretation of the results in case a detection
is made. Generally, a hierarchy of braking indices was expected, namely
$3$ for electromagnetic dipole emission, $5$ for gravitational waves
from an ellipticity of the star (or electromagnetic quadrupole emission)
and $7$ for r-mode emission%
\footnote{If several mechanisms are relevant intermediate values are possible
\cite{1983bhwd.book.....S}.%
}. Our results show that analogous to electromagnetic emission, where
the observed braking indices are below $3$, the possible braking
indices due to r-modes are likewise systematically lower and the uncertainty
in the unknown saturation mechanism is even larger - to the point
where even a discrimination between electromagnetic and gravitational
wave spindown becomes questionable. Therefore it should be hard to
determine the spindown mechanism from current electromagnetic pulsar
timing data. However, future gravitational wave observations could
changes this. Comprehensive information on the source, the gravitational
wave amplitude and the timing data, could due to the distinct r-mode
braking indices eq.~(\ref{eq:effective-braking-index}) and the quantitative
prediction of the r-mode signal eq.~(\ref{eq:independent-strain})
eventually even allow us to determine the r-mode saturation mechanism
observationally.

As discussed before, the lower boundary of the frequency range of
the expected gravitational wave signal is rather insensitive to the
details of the source and set by where the r-mode evolution exits
the instability region, see fig.~\ref{fig:instability-evo}. The
corresponding $\alpha_{{\rm sat}}$ can be obtained by equating eq.~(\ref{eq:spindown-time})
to the known age of the source which then determines the final frequency.
In contrast, the upper boundary is not restricted by gravitational
wave emission alone because the r-mode amplitude could in principle
be arbitrary small. Yet, in this case other spindown mechanisms will
dominate and set a limit on the frequency to which a potential pulsar
would have been spun down at the age of a considered source. To estimate
this effect, we assume again the spindown model eq. (\ref{eq:generic-spindown-model})
with a fixed braking index $n$. The proportionality constant is related
to the characteristic age $t_{c}\equiv-\left(f/\dot{f}\right)_{0}$
of a given pulsar and the evolution is then given by

\begin{equation}
f\left(t\right)\approx\left(\frac{1}{f_{i}^{n-1}}+\frac{\left(n-1\right)t}{t_{c}f_{0}^{n-1}}\right)^{-\frac{1}{n-1}}\:.\label{eq:generic-spindown-solution}
\end{equation}
For a conservative estimate of the upper limit we assume a braking
index of $n=2.5$ and choose the initial frequency $f_{i}$ as the
Kepler frequency of our $1.4\, M_{\odot}$ model. Fiducially choosing
the population of young stars as those in the ATNF database with spindown
rate $\dot{f}>10^{-11}s^{-2}$, the largest value for $t_{c}f_{0}^{n-1}$
is, as could have been expected, obtained for J0537-6910. Let us assume
that the observed population of pulsars is characteristic for the
spindown behavior of so far unobserved sources. This then allows us
to estimate an upper limit for the frequency to which other spindown
mechanisms would have spun down a (potentially unobserved) compact
star in a considered source even when gravitational wave emission
would be negligible for the spindown. Finally, consider pulsars with
known timing solution, where both the frequency and the spindown rate
are known. If r-mode emission dominates the current spindown, eq.~(\ref{eq:Omega-equation-saturation})
determines the saturation amplitude (see fig.~\ref{fig:F0F1data})
and in the case of J0537-6910 we find $\alpha_{{\rm sat}}\approx0.084$.
If other spindown mechanisms are important as well, $\alpha_{{\rm sat}}$
and the resulting gravitational strain would be lower.

Our results are shown in fig.~\ref{fig:strain-sources}. Again the
sources without known timing data should be compared to the thin curves,
whereas the expected strain for the pulsar J0537-6910 should be compared
to the thick curves. As can be seen there should be a realistic chance
to see Cassiopeia~A, and potentially a compact remnant in the youngest
known galactic supernova G1.9+0.3, from a coherent analysis of a month
of LIGO S6 data. Even more strikingly the frequency range where Cas~A
is above the detector sensitivity coincides with the expected frequency
range for this source. For Cas~A a detailed previous LIGO analysis
failed to detect a signal, but imposed strict bounds on a potential
r-mode amplitude of this source \cite{Abadie:2010hv}. Yet, this was
done with only 12 days of S5 data. As can be seen by the uppermost
thin detector curve in fig.~\ref{fig:strain-sources}, which shows
the S5 sensitivity for this shorter observation interval, we find
that the signal is just at the detection limit and even this only
over a narrow frequency range, so that it is not too surprising that
no detection was made. Cas A should therefore be a prime target for
advanced LIGO and according to our results it should even be promising
to perform another search of the S6 data if a coherent analysis of
a whole month of data is computationally feasible. The signal of 1987A
is about an order of magnitude smaller, but it would be in the reach
of advanced LIGO which could detect it over the entire expected frequency
range (which is larger due to its young age). The youngest candidate
pulsar, produced by SN 1957D, would be an ideal target, but it is
unfortunately too distant to detect gravitational waves due to r-mode
emission with advanced LIGO and more sensitive future detectors will
be required. Our results show that as soon as the pulsar is directly
detected and timing data becomes available this source should be in
the range of the forthcoming Einstein telescope \cite{Punturo:2010zz},
which has an order of magnitude higher sensitivity compared to advanced
LIGO. If against all odds the fastest directly observed young pulsar
J0537-6910 is still in the final stage of its r-mode evolution there
is a good chance that its signal, which was below the threshold of
the original LIGO detector, will be detected by advanced LIGO.

The relevant ranges for the timing parameters, given in table \ref{tab:source-examples},
should help to perform future searches. If r-mode emission dominates
the spindown, the frequency and spindown rate are related by eq.~(\ref{eq:Omega-equation-saturation})
and the time variation over the observation interval is given by the
semi-analytic solution eq.~(\ref{eq:spindown-solution}). In reality
where other spindown mechanisms are present as well there should nevertheless
be a strong correlation between frequency and spindown rate so that
it is not necessary to search the complete range in frequency and
spindown rate but a (narrow) parameter region around the ideal r-mode
spindown behavior should be sufficient. This could strongly reduce
the required parameter range in dedicated r-mode search. In case of
Cas~A the frequency range obtained here by our dynamical analysis
roughly agrees with the previously used fiducial range $100\,{\rm Hz}\leq\nu\leq300\,{\rm Hz}$
chosen by sensitivity arguments \cite{Abadie:2010hv}. Finally it
is important to note that the saturation amplitudes corresponding
to the above relevant gravitational frequency ranges, as denoted by
the dots in fig.~\ref{fig:strain-sources}, are precisely in the
range $10^{-2}\lesssim\alpha_{{\rm sat}}\lesssim10^{-1}$ which is
required for an explanation of the low frequencies of observed young
pulsars within the r-mode scenario discussed in section \ref{sec:Spindown-of-young}.
This should provide an incentive to search for gravitational wave
emission from r-modes of sufficiently young compact sources.

\begin{table*}
\begin{tabular}{|c|c|c|c|c|c|c|c|c|c|}
\hline 
source & age $\left[\mathrm{y}\right]$ & $D\,\left[\mathrm{kpc}\right]$ & $\nu\,\left[{\rm Hz}\right]$ & $\dot{\nu}\,\left[10^{-9}\,{\rm s^{-2}}\right]$ & $h_{0}$ & $h_{c}$ & $\left.\frac{S}{N}\right|_{\mathrm{LIGO}}^{\left(obs\right)}$ & $\left.\frac{S}{N}\right|_{\mathrm{aLIGO}}^{\left(obs\right)}$ & $\left.\frac{S}{N}\right|_{\mathrm{aLIGO}}^{\left(tot\right)}$\tabularnewline
\hline 
J0537-6910 & $?$ & $52$ & $83$ & $-0.27$ & $3.6\times10^{-26}$ & $6.5\left(6.8\right)\times10^{-20}$ & $2.1\left(0.2\right)$ & $28\left(2.8\right)$ & $-$\tabularnewline
\hline 
SN 1987A & $25$ & $51.5$ & $\left[84,992\right]$ & $\left[-18,-265\right]$ & $2.8\times10^{-25}$ & $0.8\left(1.2\right)\times10^{-19}$ & $23\left(18\right)$ & $221\left(236\right)$ & $2.0\left(3.2\right)\times10^{3}$\tabularnewline
\hline 
SN 1957D & $55$ & $4610$ & $\left[86,801\right]$ & $\left[-8.3,-155\right]$ & $2.1\times10^{-27}$ & $0.9\left(1.3\right)\times10^{-21}$ & $0.2\left(0.1\right)$ & $1.6\left(1.8\right)$ & $19\left(33\right)$\tabularnewline
\hline 
G1.9+0.3 & $\approx140$ & $7.7$ & $\left[89,545\right]$ & $\left[-3.3,-59\right]$ & $8.0\times10^{-25}$ & $4.9\left(7.2\right)\times10^{-19}$ & $57\left(62\right)$ & $598\left(662\right)$ & $0.9\left(1.8\right)\times10^{4}$\tabularnewline
\hline 
Cassiopeia A & $\approx300$ & $3.4$ & $\left[91,346\right]$ & $\left[-1.5,-19\right]$ & $1.2\times10^{-24}$ & $1.0\left(1.5\right)\times10^{-18}$ & $83\left(106\right)$ & $906\left(1015\right)$ & $1.3\left(3.7\right)\times10^{4}$\tabularnewline
\hline 
\end{tabular}

\caption{\label{tab:source-examples}Expected gravitational wave signal and
expected range of the timing parameters for different young sources
in case there is an associated fast rotating compact object that spins
down via the r-mode mechanism. The alternative sensitivity measures
in the right part of the table are discussed in detail in appendix
\ref{sec:Alternative-sensitivity-measures} and they generally strongly
overestimate the detectability since they do not take into account
our limited knowledge of the source or not even the limited observation
time. All values are given for a $1.4\, M_{\odot}$ APR star with
$\alpha_{{\rm sat}}=0.1\left(0.01\right)$; for $\alpha_{{\rm sat}}=1$
r-modes would have had already decayed at the current age of these
systems. The observable signal-to-noise ratios are computed assuming
an observation time $\Delta t=1\,{\rm y}$. The theoretical total
signal-to-noise ratios are here obtained by taking into account the
total frequency interval $\left[\nu_{f},\bar{\nu}\right]$ that is
observable from the stars current age to the end of the r-mode spindown.}
\end{table*}

\section{Conclusions}

We have derived general semi-analytic results for the r-mode spindown
evolution eqs.\ (\ref{eq:spindown-solution}) and (\ref{eq:temperature-evolution})
of young pulsars that completely reveal the dependence on the relevant
physical ingredients. We find that the key quantities like the final
spin frequency eq.\ (\ref{eq:final-frequency}) are remarkably insensitive
to the microphysical material properties. For macroscopic input quantities
like the moment of inertia of the star or the angular momentum of
the mode we gave rigorous bounds that allow us to estimate the uncertainty
in these parameters. We show that the thermal evolution is always
faster than the spindown evolution. Therefore, if the form of the
instability region is convex, a steady state is reached where viscous
heating equals cooling due to neutrino emission. Non-convexity of
the instability region due to stability windows arising from enhanced
dissipation due to exotic phases of matter can prevent the star from
reaching the thermal steady state and thereby drastically change the
evolution which could lead to clear astrophysical signatures \cite{Andersson:2001ev}.

The semi-analytic results for the final frequency of the evolution
enabled a quantitative comparison to pulsar timing data and we find
that all stars except one lie clearly below the frequency to which
r-modes can spin down a young pulsar. Only the fastest spinning young
star PSR J0537-6910 is fast enough that it could lie within the instability
region. Using additional observations of spindown rates we find that
this is unlikely since it lies just at the lower border of the uncertainty
range. The spindown time in contrast depends strongly on the saturation
amplitude and increases quickly from time scales of order years at
$\alpha_{{\rm sat}}\approx1$ to more than a million years below $\alpha_{{\rm sat}}<10^{-3}$.

The r-mode scenario can therefore provide a quantitative explanation
of the low spin frequencies of young pulsars if the saturation amplitude
is within certain limits. On the one hand it has to be sufficiently
large $\alpha_{{\rm sat}}\gtrsim0.01$ for the spindown to be fast
enough to dominate other possible spindown mechanisms and result in
spindown times less than a few hundred years, which is shorter than
the age of observed pulsars. On the other hand, even taking into
account the significant uncertainties, a very large saturation amplitude
$\alpha_{{\rm sat}}\approx1$ would have spun PSR J0537-6910 down
to its current frequency in less than a hundred years. This would
be in clear conflict with the much larger age estimate of its remnant
of roughly $5000$ years \cite{Wang:1998} and the significant present
spindown rate of the pulsar, which is most likely due to electromagnetic
radiation and should have further sped up the spindown. Different
saturation scenarios - which are here studied in a generic way via
a function $\alpha_{{\rm sat}}\!\left(\Omega,T\right)$ - do not significantly
change the spindown time and therefore should not qualitatively affect
our bounds on the saturation amplitude. In case r-modes reach sizable
amplitudes $0.01\lesssim\alpha_{{\rm sat}}\lesssim0.1$, PSR J0537-6910
could be younger than its characteristic age - obtained within a pure
magnetic dipole model - suggests. If in contrast r-modes saturate
already at amplitudes $\alpha_{{\rm sat}}\ll0.01$ they would be subleading
compared to the other spindown mechanisms, which must be sizable according
to the significant spindown rates of observed young pulsars.

Since we find that probably no observed pulsar is at present spinning
down due to r-modes, one has to be careful when interpreting spindown
limits on r-mode amplitudes from young stars \cite{Owen:2010ng}.
Although these limits give important information on these stars, they
are trivially fulfilled if the r-mode has already decayed. In this
case the current rate involves only information on other spindown
mechanisms, like magnetic dipole emission, and cannot give a bound
on the r-mode saturation amplitude during the era when r-modes were
actually present. The spindown limit \cite{Owen:2010ng} simply extrapolates
the solid lines in fig.~\ref{fig:F0F1data} outside the instability
region. In reality for a large saturation amplitude the star would
actually spin down quickly inside but then the spindown rate would
drop to a lower value determined by the other mechanisms once the
star leaves the instability region. The slope of the curves for other
spindown mechanisms are generally smaller, as illustrated in fig.~\ref{fig:F0F1data}
for the Crab and the Vela pulsar where a reliable braking index is
available. An initial r-mode spindown can therefore be perfectly consistent
with observations even for a large saturation amplitude as long as
the star is currently already sufficiently far away from the instability
boundary. Yet, in case of J0537-6910 which is still close to the instability
boundary, the spindown limit is more restrictive and coincides with
the limits given in the previous paragraph \cite{Owen:2010ng}.

We considered in this work only the dominant $m=2$ r-mode. The inclusion
of higher multipoles might slightly affect the spindown time by a
factor $\lesssim1$ but will not change the final frequency since
the corresponding instability regions of these modes are smaller \cite{Alford:2010fd}
and therefore the final segment of the steady-state curve, where the
evolution spends by far the longest time, is identical. Furthermore,
we have here mainly studied the simplest case of a neutron star with
damping due to leptonic interactions and modified Urca processes in
the core. The main reason for this is that it is the case of minimal
damping whereas other phases of matter or structural inhomogeneities
will increase the damping \cite{Jaikumar:2008kh,Alford:2010fd,Ho:2011tt,Haskell:2012}.
Therefore, the final spindown frequencies we find are the lower limit
to which general compact stars can be spun down by r-modes. In reality
several aspects like pairing, mutual friction, the crust and other
spindown mechanisms complicate the analysis. Although we did not study
these in detail here, our general semi-analytic expressions allow
us to study any single source of damping, such as the effect of an
Ekman layer.

Nuclear pairing could strongly change the temperature dependence of
material properties over the considered temperature regime so that
it would not have a simple power law form. Therefore it should alter
thermal evolution, but since at saturation the thermal evolution does
hardly affect the spindown (aside from the fact that its final frequency
at the boundary of the instability region depends on temperature)
the results presented here should not be strongly affected by pairing.
The relevant part of the boundary is determined by shear viscosity
which is dominated by electromagnetic leptonic processes which are
only indirectly affected by the presence of hadronic matter. As found
in \cite{Shternin:2008es} this is a moderate effect and taking into
account the insensitivity of our results to the shear viscosity this
should only slightly increase the considered error range. 

As discussed before mutual friction in superfluid systems is an additional
source of damping that could considerably shrink the instability region.
If this effect is very strong it could significantly raise the frequency
to which r-modes can spin down a star. As can be seen from fig.~\ref{fig:instability-evo}
if the enhanced damping is only present at low temperatures, e.g.\ $T<10^{9}$
K as assumed in \cite{Owen:1998xg}, this would affect the evolution
strongly at low saturation amplitudes but only moderately at sufficiently
large amplitudes favored above.

Furthermore, it has been argued that the boundary layer at the interface
between the core and the crust strongly enhances the viscous damping
which would reduce the r-mode instability region \cite{Lindblom:2000gu}.
This mechanism is based on the picture that there is a sharp boundary
and the fluid velocity drops to zero over a distance of a few centimeters.
However, in reality the crust is a complicated structure with possible
geometrically structured pasta phases and an inner crust where the
concentration of the neutron fluid decreases continuously. The transition
from a pure fluid to a rigid lattice then happens gradually over a
large distance and it is questionable why the above picture should
apply here. Phenomenologically these effects have partly been modeled
via a slippage parameter and the impact becomes rather small for likely
parameter values \cite{Haskell:2012}. We note, however, that our
general analytic expressions should also apply when the crust provides
the dominant source of dissipation, as encoded in the values of the
general parameters in table \ref{tab:Integral-parameters.}. In this
case further complications like the breaking or melting of the crust
could be relevant which in turn should reduce the effect of viscous
boundary layer damping.

Finally, we computed the gravitational wave emission due to the r-mode
spindown of young stars. We compare our results in terms of the intrinsic
strain amplitude to the sensitivity of the LIGO and advanced LIGO
detector for realistic searches for sources with and without timing
solution. This properly takes into account the finite observation
interval of such searches which had been neglected in initial r-mode
analyses \cite{Owen:1998xg}. We point out the remarkable property
that the intrinsic strain amplitude is \emph{independent} of the saturation
amplitude and almost independent of the saturation model which only
enters via a constant factor close to one. Therefore, the gravitational
wave signal depends effectively only on the age of the source and
its distance. The reason for this independence is that the emitted
gravitational wave signal increases both with the r-mode amplitude
and the frequency. A star of a given age will still be spinning quickly
if the saturation amplitude is low, but will have spun down to a low
frequency if the saturation amplitude is high. These competing effects
cancel each other. For realistic observation intervals, we find that
LIGO could have seen young sources in the galaxy (distance 10 kpc)
with known timing solution up to an age of about $10^{4}$ years,
whereas sources without timing solution that are older than a few
tens of years would have been undetectable. Advanced LIGO in contrast
should increase both of these age ranges by about a factor of a hundred.
Since according to our analysis none of the observed pulsars is likely
to be in the r-mode phase, very young sources without a timing solution
should be promising targets for gravitational wave searches, and advanced
LIGO should be able to detect them over the entire relevant age range
of up to thousands of years. For more distant sources in the local
group of galaxies (distance 1 Mpc) these ages are reduced by a factor
of ten thousand so that in this case probably only sources with a
known timing solution are detectable. We also considered a few explicit
examples and find that Cas A, J0537-6910, and potential compact remnants
in SN1987A and G1.9+0.3 are very promising targets that are within
the sensitivity range of advanced LIGO. To restrict the computational
cost of future searches we also provided estimates for the relevant
search ranges of the unknown timing parameters of these sources. 

The quantitative r-mode spindown scenario discussed here provides
strong evidence that young pulsars could indeed emit observable gravitational
wave signals. Both the steadily growing list of known young pulsars
and the strongly increased sensitivity of second generation gravitational
wave detectors make it a realistic possibility to detect gravitational
radiation from r-mode oscillations in the near future. Future third
generation gravitational wave detectors like the currently planned
Einstein telescope \cite{Punturo:2010zz} should then allow us to
see many more potential sources at larger distances. This would mark
the advent of gravitational wave astronomy as a way to learn about
the internal composition of compact stars.

\appendix

\section{Very low saturation amplitude scenario\label{sec:Very-low-saturation}}

The quantitative explanation of the low spin frequencies of observed
young pulsars discussed in section \ref{sec:Spindown-of-young} points
to a large saturation amplitude $\alpha_{{\rm sat}}\!>\!10^{-4}$.
In this case the expected gravitational signal is large, independent
of the saturation amplitude and even mostly of the saturation mechanism.
However, one cannot completely exclude that the striking explanation
of the pulsar spin data is a mere coincidence and the saturation amplitude
is much smaller. For completeness we therefore discuss here this (unlikely)
alternative scenario. Rather small saturation amplitudes are predicted
by mode coupling models \cite{Arras:2002dw,Bondarescu:2008qx,Bondarescu:2013xwa},
whereas other mechanisms like non-linear hydrodynamic effects or suprathermal
viscosity lead to large saturation amplitudes. In case the amplitude
is very small r-modes are irrelevant for the spindown since other
spindown mechanisms like magnetic dipole emission dominate. The spindown
evolution is then independent of the saturation amplitude. Using the
generic parametrization eq.~(\ref{eq:generic-spindown-model}) with
a braking index $n$ the spindown is given by eq.~(\ref{eq:generic-spindown-solution}).
Nevertheless these low amplitude r-modes would emit gravitational
waves. The gravitational wave strain is then given by eq.~(\ref{eq:gw-strain}),
where the frequency evolution is determined by the the appropriate
alternative spindown mechanism. For a standard neutron star spinning
with frequency $f$ we find in case of a small saturation amplitude

\[
h_{0}\approx9.0\times10^{-26}\alpha_{{\rm sat}}\left(\frac{f}{100\,{\rm Hz}}\right)\left(\frac{1\,{\rm Mpc}}{D}\right)
\]
In this case the strain is proportional to the saturation amplitude
since the spindown evolution is independent of the r-mode emission
and there is no cancellation analogous to eq.~(\ref{eq:independent-strain}).
Therefore the strain is, according to our assumption that $\alpha_{{\rm sat}}$
is small, strongly suppressed. We see that at an amplitude $\alpha_{sat}\!\approx\!10^{-4}$
there is only a chance to detect the gravitational waves with advanced
LIGO if the source spins fast and is very close. At even lower amplitudes
$\alpha_{sat}\!\ll\!10^{-4}$ a source would be undetectable.

\section{Alternative sensitivity measures\label{sec:Alternative-sensitivity-measures}}

For the analysis of the gravitational wave emission due to r-modes
a variety of measures to compare to detector sensitivities have been
used. For comparison with other work we discuss these in the following.

\subsection{Power spectrum}

An alternative way to discuss r-mode emission is via the Fourier spectrum.
In principle this is interesting for periodic sources like neutron
stars since the frequency hardly changes over the short observation
interval. The Fourier-transformed gravitational wave strain $\tilde{h}\left(\nu\right)$
in the frequency domain can be obtained in a stationary phase approximation
as

\begin{equation}
\left|\tilde{h}\!\left(\nu\right)\right|^{2}=\left|h\!\left(t\right)\right|^{2}\left|\frac{dt}{d\nu}\right|\:.\label{eq:h-tilde-connection}
\end{equation}
As seen from fig.~\ref{fig:gravitational-wave-strain}, there are
two qualitatively different stages of the gravitational wave emission.
The first arises from the growth phase of the r-mode and results in
narrow spike in $\tilde{h}\!\left(\nu\right)$ at the stars initial
frequency. This phase lasts only for a short time of order minutes
and as shown in \cite{Owen:1998xg} there is not enough energy in
this pulse to be detectable. Therefore, we neglect this initial part
of the spectrum here. In the subsequent saturated phase where the
star spins down it emits gravitational waves over a continuous frequency
range with strain density

\begin{equation}
\tilde{h}\!\left(\nu\right)=\sqrt{\frac{9\tilde{I}GMR^{2}}{20D^{2}\nu}}\:.\label{eq:h-tilde-saturation}
\end{equation}
The quantity that is usually used for a comparison with the detector
sensitivity is the characteristic amplitude of the signal defined
by $h_{c}\equiv\nu\,\tilde{h}$. In particular the minimum characteristic
amplitude $h_{c,f}\equiv h_{c}\!(\nu_{f})$ reached at the lower frequency
boundary of the evolution is interesting. The square root in eq.~(\ref{eq:h-tilde-saturation})
makes the dependence of $h_{c,f}$ on the microscopic parameters even
weaker than for the final frequency eq.~(\ref{eq:final-frequency-NS}).
However, the expression depends more strongly on macroscopic quantities
like the mass and the moment of inertia encoded in the parameter $\tilde{I}$.
The rigorous bounds eq.~(\ref{eq:I-tilde-bounds}) show that this
dependence is also moderate since all these quantities vary only within
roughly a factor of two. For a standard neutron star we find

\begin{align}
 & h_{c,f}^{\left(NS\right)}\approx1.6\cdot10^{-22}\left(\!\frac{\tilde{S}}{\tilde{S}_{{\rm fid}}}\!\right)^{\!3/46}\!\!\left(\!\frac{\tilde{L}}{\tilde{L}_{{\rm fid}}}\!\right)^{\!5/368}\!\!\left(\!\frac{\tilde{J}}{\tilde{J}_{{\rm fid}}}\!\right)^{\!-29/184}\nonumber \\
 & \times\!\left(\!\frac{\tilde{I}}{\tilde{I}_{{\rm fid}}}\!\right)^{\!1/2}\!\!\left(\!\frac{M}{1.4\, M_{\odot}}\!\right)^{\!63/184}\!\!\left(\!\frac{R}{11.5\,\mathrm{km}}\!\right)^{\!281/368}\!\!\left(\!\frac{\mathrm{Mpc}}{D}\!\right)\alpha_{{\rm sat}}^{-5/184}\,.\label{eq:minimum-characteristic-amp-NS}
\end{align}
Note that this quantity is also nearly independent of the saturation
amplitude $\alpha_{{\rm sat}}$ (and even increases with decreasing
amplitude). The same holds for the microscopic parameters. For example
the dependence on the neutrino emissivity involves the power $5/368$
which provides a striking example of how insensitive such macroscopic
observables can be to microscopically decisive yet inherently poorly
known quantities, since even a drastic change by three orders of magnitude
would only have an irrelevant effect of less than $10\%$. 

The detector sensitivity is described by the rms strain noise $h_{{\rm rms}}\!\equiv\!\sqrt{\nu S_{h}\!\left(\nu\right)}$
in terms of the power spectral density $S_{h}$ of the detector noise.
In contrast to eq.~(\ref{eq:observational-intrinsic-strain}) this
quantity is only a measure of the general sensitivity of the detector
but not of a particular search and contains neither information on
the source nor on the observation interval. In order for the characteristic
amplitude to give a meaningful estimate the observation time should
be comparable to the time scale of the frequency change of the signal
(so that all strength of a given frequency is actually detected).
Clearly for small saturation amplitudes where the spindown time increases
drastically with decreasing $\alpha_{{\rm sat}}$, see eq.~(\ref{eq:spindown-time-NS})
and table~\ref{tab:results}, this is not guaranteed and therefore
the fact that the characteristic amplitude is above the background
alone is not sufficient for a detection.

\begin{figure}
\includegraphics[scale=0.85]{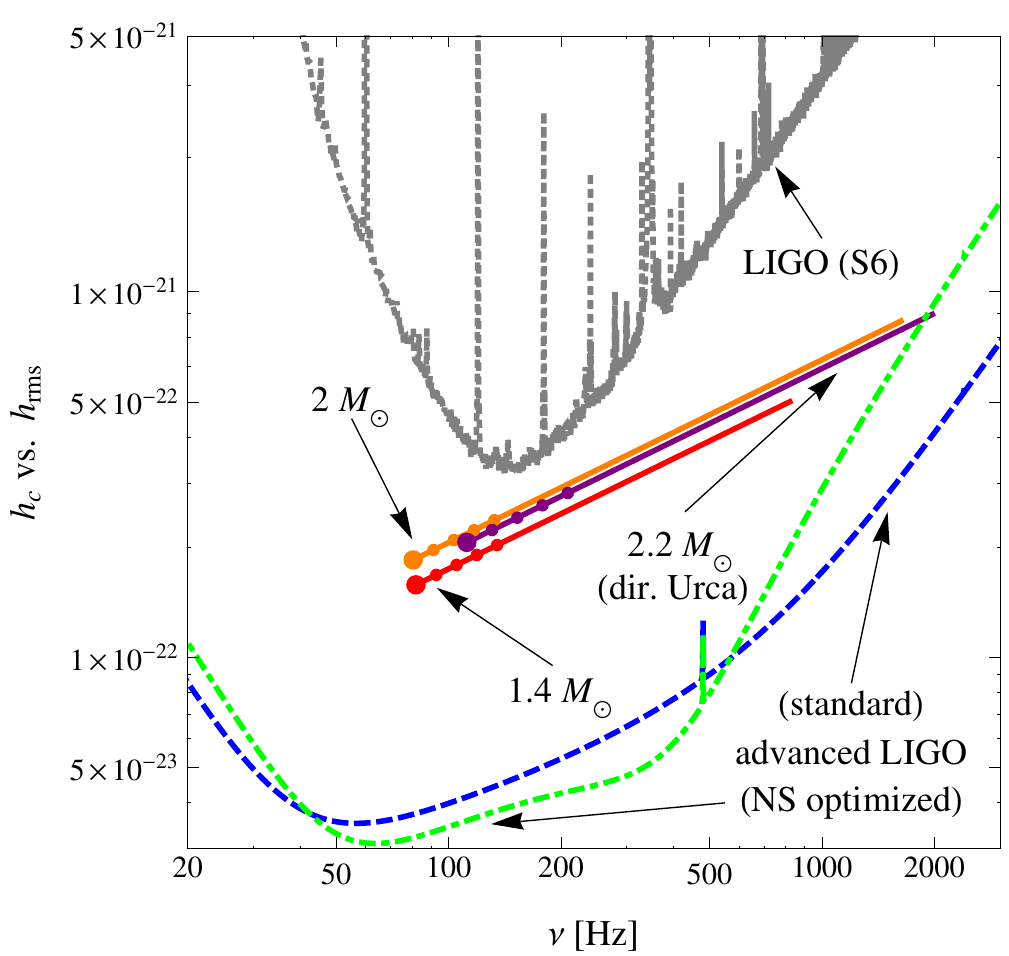}

\caption{\label{fig:detectability}The characteristic gravitational wave amplitude
of neutron stars, located somewhere in the Virgo cluster (distance
20 Mpc) and based on the different APR stars given in tab.~\ref{tab:parameter-values},
compared to the sensitivity of the LIGO (dotted) and advanced LIGO
(dashed) \cite{Harry:2010zz} detectors. The large dots denote the
analytic expressions for the lower frequency limit for an amplitude
$\alpha_{{\rm sat}}=1$ and the smaller ones for lower amplitudes
down to $\alpha_{{\rm sat}}=10^{-4}$.}
\end{figure}

The result for the characteristic amplitude in the saturated regime
is given for different stars in fig.~\ref{fig:detectability}. Whereas
the spectra of the two standard neutron stars with modified Urca reactions
have nearly the same lower frequency cutoff, the heavy star with direct
Urca interactions in the core has a higher cutoff $>100$ Hz, but
otherwise features a characteristic amplitude of similar size, as
could have been expected from the insensitivity to the neutrino emissivity
in eq.~(\ref{eq:final-frequency}).

\subsection{Theoretical signal detectability}

In \cite{Owen:1998xg} the following expression for the signal to
noise ratio for matched filtering was given 

\begin{equation}
\left.\left(\frac{S}{N}\right)^{\!2}\right|^{\left({\rm tot.}\right)}\!=\!2\int_{\nu_{min}}^{\nu_{max}}\!\frac{d\nu}{\nu}\left(\!\frac{h_{c}}{h_{{\rm rms}}}\!\right)^{\!2}\!=\!\frac{9I}{10D^{2}}\int_{\nu_{min}}^{\nu_{max}}\!\!\frac{d\nu}{\nu\, S_{h}\!\left(\nu\right)}\label{eq:signal-to-noise-integral}
\end{equation}
in terms of the power spectral density $S_{h}$ of the noise and ranging
over a range from the final frequency of the r-mode evolution $\nu_{min}=\nu_{f}$
to the initial frequency $\nu_{max}=\nu_{i}$ at which the star is
created. Here the label $\left({\rm tot.}\right)$ has been added
to indicate that the total frequency spectrum is included in this
quantity. This point will be discussed in more detail below. In eq.~(\ref{eq:signal-to-noise-integral})
logarithmic frequency intervals are weighted by $1/S_{h}$, since
at low frequencies the frequency changes slowly which makes it easier
to detect the periodic signal. After a numerical integration an
approximate fit to the result over the relevant parameter ranges for
$\nu_{min}$ and $\nu_{max}$ is given for the advanced LIGO detector
by%
\footnote{We fit the integrated form rather than the initial strain noise $S_{h}$,
since this provides a more precise result. Fitting the strain noise
with a double power law $S_{h}\left(\nu\right)\approx a\nu^{-2}+b\nu^{2}+c$,
similar to the ansatz used in \cite{Owen:2010ng} before their subsequent
approximation, we can perform the integration analytically, yielding
the same qualitative leading parameter dependence on $\nu_{min}$
(power law) and $\nu_{max}$ (logarithmic) we use for the fit in eq.~(\ref{eq:SN-aLIGO}).%
}

\begin{align}
 & \left.\frac{S}{N}\right|_{\mathrm{aLIGO}}^{\left({\rm tot.}\right)}\approx4.09\cdot10^{23}\sqrt{{\rm Hz}}\frac{\sqrt{GI}}{D}\label{eq:SN-aLIGO}\\
 & \quad\;\times\sqrt{1+0.40\log\left(\frac{\nu_{max}}{1000\,{\rm Hz}}\right)+0.36\frac{100\,{\rm Hz}-v_{min}}{100\,{\rm Hz}}}\;.\nonumber 
\end{align}
Compared to the result given in \cite{Owen:2010ng} for the anticipated
second generation LIGO detector we find a much weaker dependence on
the minimum final frequency and a stronger dependence on the unknown
maximum initial frequency. By inserting the result for the minimum
frequency eq.~(\ref{eq:final-frequency-NS}) we see that the uncertainty
of this result due to the microphysics is minor. The main uncertainty
arises from the moment of inertia which using the above bounds has
a maximum theoretical uncertainty given in eq.~(\ref{eq:moment-of-inertia-bounds}).

The signal to noise ratios for different sources at different distances
are shown in table~\ref{tab:semi-analytic-results}. Here we compare
the values for sources given by the different considered stars listed
in tab.~\ref{tab:parameter-values}. These are assumed to be located
at different distances, namely in the Virgo Cluster (20 Mpc), the
Local group of galaxies (1 Mpc) and within the Milky way (30 kpc).
Using the actual anticipated noise background of the advanced LIGO
detector and without the assumption that the evolution stops at $10^{9}$
K, our values are slightly larger than those obtained in \cite{Owen:1998xg}.
Heavier stars tend to give larger values due to the larger moment
of inertia whereas direct Urca processes reduce the signal to noise
ratio since the thermal steady-state curve is at lower temperatures
and its intersection with the boundary of the instability region is
correspondingly at a higher frequency that sets the lower gravitational
wave frequency cutoff in eq.~(\ref{eq:SN-aLIGO}). Yet, overall these
values are rather similar for the different considered stars. These
considerations suggest that theoretically a detection is possible
for sources up to large distances in case the gravitational wave signal
could be detected over its complete frequency range. However, this
is inherently impossible for a continuous r-mode signal.

\begin{table*}
\begin{tabular}{|c|c|c|c|c|c|c|c|c|c|c|c|c|}
\hline 
neutron star & $T_{f}\left[\mathrm{K}\right]$ & $f_{f}\left[\mathrm{Hz}\right]$ & $\nu_{f}\left[\mathrm{Hz}\right]$ & $t_{sd}\left[y\right]$ & $h_{f}$ @ $1$ Mpc & $h_{c,f}$ @ $1$ Mpc & $\left.\frac{S}{N}\right|_{\mathrm{Virgo}}^{\left(tot\right)}$ & $\left.\frac{S}{N}\right|_{\mathrm{L.\, g.}}^{\left(tot\right)}$ & $\left.\frac{S}{N}\right|_{\mathrm{M.\, w.}}^{\left(tot\right)}$ & $\left.\frac{S}{N}\right|_{\mathrm{Virgo}}^{\left(obs\right)}$ & $\left.\frac{S}{N}\right|_{\mathrm{L.\, g.}}^{\left(obs\right)}$ & $\left.\frac{S}{N}\right|_{\mathrm{M.\, w.}}^{\left(obs\right)}$\tabularnewline
\hline 
APR $1.4\, M_{\odot}$ & $1.26\cdot10^{9}$ & $61.4$ & $81.8$ & $12.3$ & $7.2\cdot10^{-27}$ & $1.6\cdot10^{-22}$ & $11.7$ & $234$ & $7794$ & $0.084$ & $1.7$ & $56$\tabularnewline
\hline 
APR $2.0\, M_{\odot}$ & $1.39\cdot10^{9}$ & $60.3$ & $80.4$ & $4.9$ & $9.7\cdot10^{-27}$ & $1.8\cdot10^{-22}$ & $14.0$ & $281$ & $9356$ & $0.099$ & $2.0$ & $66$\tabularnewline
\hline 
APR $2.21\, M_{\odot}$ & $3.15\cdot10^{8}$ & $84.0$ & $112$ & $1.7$ & $2.2\cdot10^{-26}$ & $2.1\cdot10^{-22}$ & $12.7$ & $254$ & $8460$ & $0.094$ & $1.9$ & $62$\tabularnewline
\hline 
\end{tabular}

\caption{\label{tab:semi-analytic-results}Results for the spindown and gravitational
wave observables for different neutron stars considered in this work
obtained from the semi-analytic expressions eqs.~(\ref{eq:final-temperature}),
(\ref{eq:final-frequency}), (\ref{eq:spindown-time}), (\ref{eq:minimum-strain-NS}),
(\ref{eq:minimum-characteristic-amp-NS}) and (\ref{eq:signal-to-noise-integral}).
In contrast to the lower mass stars, the $2.21\, M_{\odot}$ neutron
star is dense enough to allow for direct Urca reactions. All values
are given for a large saturation amplitude $\alpha_{sat}=1$. The
signal to noise ratios are given for the advanced LIGO detector and
are obtained with the gravitational wave frequency corresponding to
the Kepler frequency as an upper cutoff of the spectrum respectively
a fiducial observation interval $\Delta t/t=10^{-3}$. They are given
for sources at three different distances ranging from the Virgo cluster
(20 Mpc), and the local group of galaxies (1 Mpc) to a source within
the Milky Way (30 kpc). }
\end{table*}

\subsection{Practical signal detectability}

The above total signal to noise ratios do not reflect the detectability
of the signal of a given single source over the actual observation
period, as noted already in \cite{Owen:1998xg}. They would provide
such an actual measure for the detectability only for a short multi-wavelength
signal that is completely observed, for example a binary inspiral.
However, these ratios cannot give a reasonable measure in the present
case where the source is monochromatic and changes slowly with time
while it is only observed for a short time interval, since the frequency
ranges in eq.~(\ref{eq:signal-to-noise-integral}) that would only
be detectable hundreds of years before or after the present observation
can hardly affect the detectability within the limited observation
interval. Therefore a proper signal to noise ratio that measures the
detectability must take into account the finite duration of the observation
and only include the relevant frequencies in eq.~(\ref{eq:signal-to-noise-integral}).

To find a more realistic signal to noise ratio for the detectability
in gravitational wave detectors, assume that the frequency of the
source changes only slightly during the short observation time interval
$\Delta t$ over which it has an average frequency $\bar{\nu}$. The
change $\Delta\nu\ll\bar{\nu}$ during this time interval can be obtained
from eq.~(\ref{eq:Omega-equation-saturation})

\begin{equation}
\Delta\nu\approx\frac{2}{3\pi}\frac{d\Omega}{dt}\Delta t=\frac{2\bar{\nu}Q\alpha_{{\rm sat}}^{2}}{\tau_{G}\left(\bar{\Omega}\right)}\Delta t\:.\label{eq:observed-frequency-range}
\end{equation}
The current angular velocity $\bar{\Omega}$ can be obtained in terms
of the current age $t$ of the star from the solution of the spindown
equation eq.~(\ref{eq:Omega-equation-saturation}) which yields in
the limit $\Omega_{i}\gg\bar{\Omega}$

\begin{equation}
\Delta\nu\approx-\frac{\bar{\nu}}{6}\frac{\Delta t}{t}\:.
\end{equation}
For $\Delta\nu\ll\bar{\nu}$ the integral in eq.~(\ref{eq:signal-to-noise-integral})
gives an estimate for the actual \emph{observable} signal to noise
ratio

\begin{align}
\left.\frac{S}{N}\right|^{\left({\rm obs.}\right)} & \approx\sqrt{\frac{9\tilde{I}GMR^{2}}{10D^{2}}\frac{\Delta\nu}{\bar{\nu}S_{h}\!\left(\bar{\nu}\right)}}=\sqrt{\frac{3GI}{20S_{h}\!\left(\bar{\nu}\right)D^{2}}\frac{\Delta t}{t}}\label{eq:observable-signal-to-noise}
\end{align}
Analogous to the equivalent intrinsic strain eq.~(\ref{eq:observational-intrinsic-strain}),
this quantity depends on the square of the observation time interval
$\Delta t$ (see also \cite{Watts:2008qw}) and depends on the internal
composition of the star only via the moment of inertia with uncertainty
range eq.~(\ref{eq:moment-of-inertia-bounds}). Moreover, it depends
only on the present (averaged) gravitational wave frequency $\bar{\nu}$
and not the complete spectrum. It is again not directly dependent
on the saturation amplitude which only enters indirectly via the dependence
on the current frequency via $S_{h}\!\left(\bar{\nu}\right)$. Yet,
we stress again that this expression is only valid during the later
evolution where the spindown becomes sufficiently slow so that $\Delta\nu\ll\bar{\nu}$.
The right part of table ~\ref{tab:semi-analytic-results} shows the
values for the advanced LIGO detector for the different stars and
a ratio $\Delta t/t=10^{-3}$ compared to the previous signal-to-noise
estimates. As can be seen in this case the observable signal to noise
ratios are more than two orders of magnitude smaller than the total
expressions eq.~(\ref{eq:SN-aLIGO}). Nevertheless eq.~(\ref{eq:observable-signal-to-noise})
still greatly overestimates the detectability since it does not take
into account our limited information about the source.

For the advanced LIGO detector the background is nearly constant over
the relevant range from 80 to 800 Hz with $S_{h}^{1/2}\approx4\cdot10^{-24}$
and thereby allows us to further approximate to obtain the simple
estimate

\begin{equation}
\left.\frac{S}{N}\right|_{\mathrm{aLIGO}}^{\left({\rm obs.}\right)}\approx50\,\sqrt{\frac{\Delta t}{t}}\left(\frac{1\,\mathrm{Mpc}}{D}\right)\:.\label{eq:observable-SN-aLIGO}
\end{equation}
Although this expression overestimates the detectability under realistic
conditions it might be useful for simple estimates. E.g.\ for a one
year observation period, comparable to past LIGO runs, a source in
the Virgo cluster ($20$ Mpc) would, in contrast to the early estimates
in \cite{Owen:2010ng}, be impossible to detect unless it is extraordinarily
young. A source in our local group of galaxies ($1$ Mpc) in contrast
might be detectable if the results obtained assuming matched filtering
do not strongly overestimate the detectability for more realistic
search methods. A source in the Milky Way ($30$ kpc) in contrast
could be detectable even if a more realistic search strategy has a
significantly lower signal to noise ratio. As discussed before, since
the original LIGO detector had a more than an order of magnitude larger
background, in contrast to previous estimates \cite{Owen:1998xg},
it is not surprising that no signal has been detected so far.

Finally let us estimate likely values for the above signal to noise
ratios for the examples of possible sources that have been discussed
in the main text. As can be seen in table \ref{tab:source-examples},
despite the fact that the observable signal-to-noise ratio eq.~(\ref{eq:observable-SN-aLIGO})
is reduced by more than an order of magnitude compared to the total
versions eq.~(\ref{eq:SN-aLIGO}), the values are still significantly
larger than what one would expect from the detailed analysis in terms
of $h_{0}$ in the main text. Although the relative sizes for the
individual sources are similar, the sensitivity (eq.~(\ref{eq:observational-intrinsic-strain}))
which properly takes into account our ignorance of the details of
the source leads to systematically lower sensitivities. Correspondingly,
one should generally be careful with conclusions based on these alternative
sensitivity estimates.
\begin{acknowledgments}
We are grateful to Brynmor Haskell, Wynn Ho, Prashant Jaikumar, Feryal
\"Ozel and Simin Mahmoodifar for helpful discussions and to the referee
for exceptionally constructive advice. This research was supported
in part by the Offices of Nuclear Physics and High Energy Physics
of the U.S. Department of Energy under contracts \#DE-FG02-91ER40628,
\#DE-FG02-05ER41375.
\end{acknowledgments}
\bibliographystyle{h-physrev}
\bibliography{cs.bib}

\end{document}